\documentclass[draftcls,11pt,onecolumn]{IEEEtran}

\usepackage{cite}
\usepackage{graphicx}
\usepackage[cmex10]{amsmath}
\usepackage{amsfonts,amssymb}
\usepackage{latexsym}
\usepackage{algorithmic}
\usepackage{algorithm}
\usepackage{array}
\usepackage{wrapfig}
\usepackage{mdwmath}
\usepackage{mdwtab}
\usepackage{eqparbox}
\usepackage{color}
\usepackage{enumerate}
\usepackage[utf8]{inputenc}
\usepackage{times}
\usepackage{booktabs}
\usepackage{qtree}
\usepackage[justification=centering]{caption}
\usepackage{tikz}
\usetikzlibrary{positioning}
\usepackage{mdframed}
\usepackage{showexpl}
\usetikzlibrary{decorations.pathreplacing}
\usepackage{ragged2e}
\usepackage{lettrine}
\usetikzlibrary{trees}
\usepackage{multirow,bigdelim}
\usepackage{enumitem}
\usepackage{subcaption}
\usepackage{lipsum}
\usepackage{wrapfig}

\newtheorem{example}{Example}

\newcommand{\erfc}{\mathrm{erfc}\,}
\newcommand{\ith}{i^{th}}

\newcommand{\set}{$\mathcal{P}$}

\newcommand{\displacementv}{\vec{r}}

\newcommand{\prt}{p(r,t|r_0)}
\newcommand{\prtzero}{p(r,t \rightarrow 0 |r_0)}
\newcommand{\prtat}[1]{p(#1,t|r_0)}

\newcommand{\rrn}{r_r}

\newcommand{\SqDortPiDeT}{\sqrt{4\pi D t}}
\newcommand{\fhit}{f_{\text{hit}}(t)}
\newcommand{\Fhit}{F_{\text{hit}}(t)}
\newcommand{\fhitt}[1]{f_{\text{hit}}(#1)}
\newcommand{\Fhitt}[1]{F_{\text{hit}}(#1)}
\newcommand{\ts}{t_s}

\newcommand{\argmin}{\arg\!\min}

\newcommand{\gauss}{\mathcal{N}({\mu^{\{i\}},\sigma^{2\{i\}}})}

\newcommand{\bt}[1]{bit-${#1}$}
\newcommand{\E}{\mathrm{E}}
\newcommand{\C}{C_{i}}
\newcommand{\Var}{\mathrm{Var}}

\begin{document}

\title{ISI Mitigation Techniques in Molecular Communication}

\author{
\textit{Burcu Tepekule$^{1}$, Ali E. Pusane$^{1}$, H. Birkan Yilmaz$^{2}$, Chan-Byoung Chae$^{3}$, and Tuna Tugcu$^{4}$}\\
$^{1}$ Department of Electrical and Electronics Engineering, Bogazici University,  Istanbul, Turkey\\
$^{2}$ Yonsei Institute of Convergence Technology, Yonsei University, Seoul,  Korea\\
$^{3}$ School of Integrated Technology, Yonsei University, Seoul,  Korea\\
$^{4}$ NETLAB, Department of Computer Engineering, Bogazici University, Istanbul, Turkey\\
\{burcu.tepekule, ali.pusane, tugcu\}@boun.edu.tr, 
\{birkan.yilmaz, cbchae\}@yonsei.ac.kr
}

\maketitle
\begin{abstract}

Molecular communication is a new field of communication where molecules are used to transfer information. Among the proposed methods, molecular communication via diffusion (MCvD) is particularly effective. One of the main challenges in MCvD is the intersymbol interference (ISI), which inhibits communication at high data rates. Furthermore, at the nano scale, energy efficiency becomes an essential problem. Before addressing these problems, a pre-determined threshold for the received signal must be calculated to make a decision. In this paper, an analytical technique is proposed to determine the optimum threshold, whereas in the literature, these thresholds are generally calculated empirically. Since the main goal of this paper is to build an MCvD system suitable for operating at high data rates without sacrificing quality, new modulation and filtering techniques are proposed to decrease the effects of ISI and enhance energy efficiency. As a transmitter-based solution, a modulation technique for MCvD, molecular transition shift keying (MTSK), is proposed in order to increase the data rate via suppressing the ISI. Furthermore, for energy efficiency, a power adjustment technique that utilizes the residual molecules is proposed. Finally, as a receiver-based solution, a new energy efficient decision feedback filter (DFF) is proposed as a substitute for the decoders such as minimum mean squared error (MMSE) and decision feedback equalizer (DFE). The error performance of DFF and MMSE equalizers are compared in terms of bit error rates, and it is concluded that DFF may be more advantageous when energy efficiency is concerned, due to its lower computational complexity. 
\end{abstract}

\newpage


\IEEEpeerreviewmaketitle

\section{Introduction}

Nanotechnology enables the miniaturization and fabrication of devices in a scale ranging from 1 to 100 nanometers. At this scale, a nano-machine or a nano-enabled node (NeN) can be considered as the most basic functional unit~\cite{Akyildiz20082260}. NeNs are tiny components consisting of an arranged set of molecules, which are able to perform very simple computation, sensing, and/or actuation tasks~\cite{Suda05exploratoryresearch}. They can be interconnected to form a nanonetwork, in which they can coordinate, share, and fuse information. At such small dimensions, electromagnetic communication is challenging because of physical implementation constraints, such as the ratio of the antenna size to the wavelength of the electromagnetic signal~\cite{farsad2013tabletop}. Molecular communication is a new field of communication suitable for nanonetworks, where instead of electric currents or electromagnetic waves, patterns of molecules are used to transfer information from a source (transmitter) to a destination (receiver)~\cite{farsad2012information}. In the literature, various molecular communication systems, such as molecular communication via diffusion (MCvD), calcium signaling, microtubules, pheromone signaling, and bacterium-based communication are proposed ~\cite{akyildiz2011nanonetworksAN},~\cite{6208883}. Among these systems, MCvD is a particularly effective and energy efficient method for transporting information ~\cite{kuran2010energyMF},~\cite{kim2013novelMT}. 

An MCvD system is composed of five main processes: encoding, emission (transmission), propagation, absorption (reception), and decoding. \cite{farsad2014channelAN}. In the encoding stage, the transmitter encodes the information onto a physical property (e.g., number, type, etc.) of the messenger molecules. These molecules propagate through the environment following the physical characteristics of the channel, and when some of these molecules arrive at the receiver (i.e., hit the receiver), they are sensed and absorbed by the receptors on the surface of the receiver. The properties of these received molecules constitute the received signal~\cite{6623387}, and the received signal is decoded according to the encoding technique. Communicating NeN pairs are assumed to be synchronized and the overall communication time is divided into time slots of equal duration that allow a single symbol to be sent. These time slots are called symbol durations and denoted by ${t_s}$. 

Due to the nature of diffusion, some of the messenger molecules may fail at arriving at the receiver in their intended time slots and interfere with the messenger molecules of subsequent transmissions, causing inter symbol interference (ISI). One of the most popular solutions to reduce the amount of ISI at the receiver is to keep the symbol duration as long as possible and, thus, allow the messenger molecules a longer time to reach their destinations. This effectively reduces the number of residual molecules left in the channel. On the other hand, increasing the symbol duration also decreases the data rate, which is already slow enough due to the nature of the diffusion. Additionally, another major constraint on communication at nano scale is the energy efficiency due to the extremely small size of the nano scaled devices~\cite{6476072}. Therefore, a trade off is observed between the data rate, energy efficiency, and communication quality. The main goal of this paper is to propose new techniques applicable at both transmitter and receiver sides, which will improve the overall communication quality to achieve arbitrarily low error rates at shorter symbol durations, hence increasing the data rate. 

One of the open problems in molecular communication is the lack of diversity in modulation techniques suitable for operating at high data rates efficiently. In the literature, concentration shift keying (CSK) and molecule shift keying (MoSK) are the most commonly used modulation techniques for nanonetworks where communication is realized via diffusion~\cite{5962989}. In binary CSK (BCSK), number of the received messenger molecules is used as the amplitude of the signal. The receiver decodes the intended symbol as a \bt{1} if the number of messenger molecules arriving at the receiver during a symbol duration exceeds a pre-determined threshold, and as a \bt{0}, otherwise. The binary MoSK (BMoSK), on the other hand, utilizes the emission of two different types of messenger molecules, where the transmitter releases the appropriate type of molecule based on the current symbol. The receiver then decodes the intended symbol based on the type and number of the molecules received during a time slot~\cite{5962989}. None of these modulation techniques aim to mitigate the effects of ISI directly, hence they require large signal powers to operate at low error rates, and are insufficient in terms of energy efficiency.

Two different approaches can be considered for dealing with the energy efficiency problem in a communication system. The first approach is to reduce the signal power as much as possible. In this paper, we propose a new modulation technique, molecular transition shift keying (MTSK), which is an energy efficient modulation technique designed to reduce the effects of ISI for a single-transmitter single-receiver communication system. To enhance the energy efficiency, a power adjustment technique that utilizes the residual molecules in the channel is also proposed.

Power consumption can also be decreased by reducing the power expended by a NeN during the encoding/decoding processes, which requires the design of filters or equalizers with minimum computational complexity possible. In this paper, we propose a decision feedback filter for the decoding process, which has a lower computational complexity compared to a minimum mean squared error or decision feedback equalizer. Analyzing the transmitter and receiver based ISI mitigation techniques including encoding and filtering techniques in a comprehensive manner can also be considered as one of the main contributions of this paper.
 
Another open problem in the literature is the thresholding problem. Notice that BCSK, BMoSK, and MTSK require a pre-determined threshold at the receiver to make a decision for the received signal. In the literature, these threshold values are calculated empirically, which involves calculating the bit error rate for various detection threshold values and choosing the threshold that minimizes the error rate ~\cite{kuran2010energyMF},~\cite{5962989}. This means that a long sequence of pilot symbols must be used before the information transmission in order to obtain a comprehensive understanding of the system behavior. Furthermore, even if one of the system parameters, such as temperature, diffusion coefficient, transmitter - receiver distance, etc., changes slightly, this empirical calculation must be repeated. In this paper, an analytical technique is proposed to determine the optimum threshold value prior to information transmission, which is also one of the main contributions of this paper.

The remainder of this paper is organized as follows: Section II reviews the characteristics of the diffusion process, modulation techniques in the literature, and the ISI problem. The proposed analytical technique to determine the optimum threshold value for a nano communication system is presented in Section III. Section IV introduces the transmitter based ISI mitigation techniques, which include the proposed modulation technique, MTSK, and the power adjustment method applied to different modulation techniques. Receiver based ISI mitigation techniques that include the proposed DFF and the MMSE equalizer are given in Section V. Section VI concludes the paper.

\section{Molecular communication via diffusion and ISI}

\begin{figure}[t]
\centering{\includegraphics[width=0.7\columnwidth,keepaspectratio]
{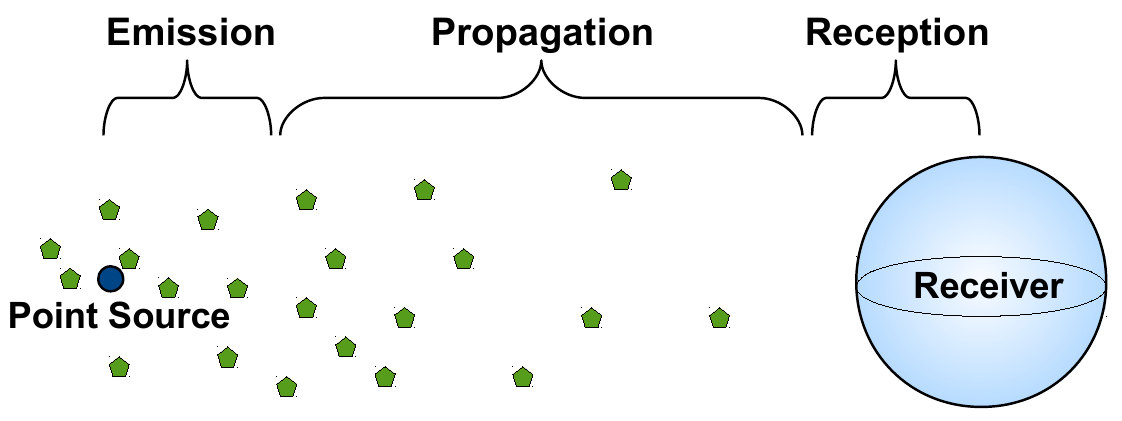}}
\caption{MCvD system model including a point source and a spherical receiver.}
\label{fig:communication_model}
\end{figure}
The communication model used in this paper is depicted in Figure  \ref{fig:communication_model}. Messenger molecules are used as information carriers between a point source and a spherical receiver with absorbing receptors. The point source is located at a distance $r_0$ from the center of the receiver. The point source and the spherical receiver both reside in a fluid propagation medium, which is a 3-dimensional (3-D) environment. After the information is modulated onto some physical property of the messenger molecules, these molecules are released to the medium, where they diffuse according to Brownian motion and arrive at the receiver. To receive the molecules (i.e., the signal), the spherical receiver with radius $r_r$, uses receptors placed on its surface.

The messenger molecules are the information particles for an MCvD system. At this scale, random movement/diffusion of particles through the fluid is modeled by Brownian motion. The motion is governed by the combined forces applied to a messenger molecule by the molecules of the fluid due to thermal energy. Brownian motion is described by the Wiener process, which is a continuous-time stochastic process. The Wiener process $W_t$ is characterized by four properties:
\begin{itemize}
\item $W_0 = 0$,
\item $W_t$ is almost surely continuous,
\item $W_t$ has independent increments,
\item $W_t - W_s \sim \mathcal{N}(0, t-s)$ for $0\leq s \leq t$.
\end{itemize}
Here, $\mathcal{N}(\mu, \sigma^2)$ denotes the Gaussian distribution with mean $\mu$ and variance $\sigma^2$. For simulating the Brownian motion in an $n$-dimensional space, time is divided into very small steps, and at each time step, random movement is applied to each dimension as
\begin{equation}
\displacementv_{t+\Delta t} = \displacementv_{t} + \Delta\displacementv .
\end{equation}
The total displacement of a molecule ($\Delta\displacementv$) in one time step  ($\Delta t$) can be found as 
\begin{equation}
\Delta\displacementv = (\Delta x_1, ... , \Delta x_n),
\end{equation}
where $\Delta x_i$ is the displacement of a molecule in the $\ith$ dimension. Movement in each dimension for a given time step is modeled independently and follows a Gaussian distribution, i.e., $\Delta x_i \sim \mathcal{N}(0, 2D\Delta t)$ $\forall i \in \{1,...,n\}$.
Molecules propagate in the environment according to these dynamics. As conventionally done in the literature, our model ignores, for simplicity, collisions between the messenger molecules~\cite{moore2009molecularCM, saxton2007modeling2A}. This model utilizing the Brownian motion is used for Monte Carlo simulations.
\subsection{Absorption rate of a perfectly absorbing spherical receiver} \label{subsec:spherical_absorber}
The microscopic theory of diffusion can be developed from the assumption that a substance will move down its concentration gradient. The derivative of the flux with respect to time results in Fick's Second Law in a 3-D environment, given by
\begin{equation}
\label{eq:ficks3}
\frac{\partial \prt}{\partial t} = D \nabla^2 \prt,
\end{equation}
\noindent where $\nabla^2$, $\prt$, and $D$ are the Laplacian operator, the molecule distribution function at time $t$ and distance $r$ given the initial distance $r_0$, and the diffusion constant, respectively. The value of $D$ depends on the temperature, the viscosity of the fluid, and the Stokes' radius of the molecule \cite{tyrrell1984diffusionIL}. 

Fraction of hitting molecules to a perfectly absorbing spherical receiver located at $(0,0,0)$ has been recently derived in \cite{yilmaz20143dChannelCF} by solving the Fick's diffusion equation with relevant initial and boundary conditions and describing the absorbing process. The initial condition is defined as
\begin{equation}
\prtzero = \frac{1}{4 \pi r_0^2} \delta(r - r_0).
\end{equation}
The first boundary condition is
\begin{equation}
\label{eq:boundary_inf}
\lim_{r \rightarrow \infty} \prt = 0,
\end{equation}
which reflects the assumption that the distribution of the molecules vanishes at distances far greater than $r_0$. The second boundary condition is
\begin{equation}
\label{eq:boundary_sphere}
D \frac{\partial \prt}{\partial r} = w \prt \text{ for } r = \rrn ,
\end{equation}
where $\rrn$ and $w$ denote the radius of the receiver and the rate of reaction, respectively. When the rate of reaction approaches infinity, it corresponds to the boundary condition in which every collision leads to an absorption. In this case, we consequently have a diminishing $\prt$ as $r$ approaches the surface of the absorber (i.e., $\lim_{r \rightarrow r_r}\prtat{r} = 0$).

After solving the differential equation for $w \rightarrow \infty$ for the perfectly absorbing sphere with the given boundary and initial conditions, the molecule distribution function at time $t$ and distance $r$ is obtained as 
\begin{align}
\begin{split}
\prt &=      \frac{1}{4\pi r r_0} \frac{1}{\SqDortPiDeT}  \left( e^{- \frac{(r-r_0)^2}{4Dt}} - e^{- \frac{(r + r_0 - 2\rrn)^2}{4Dt}} \right). \label{eq:di}
\end{split}
\end{align}

\begin{figure}
\centering
  \centering
  \includegraphics[width=0.7\linewidth]{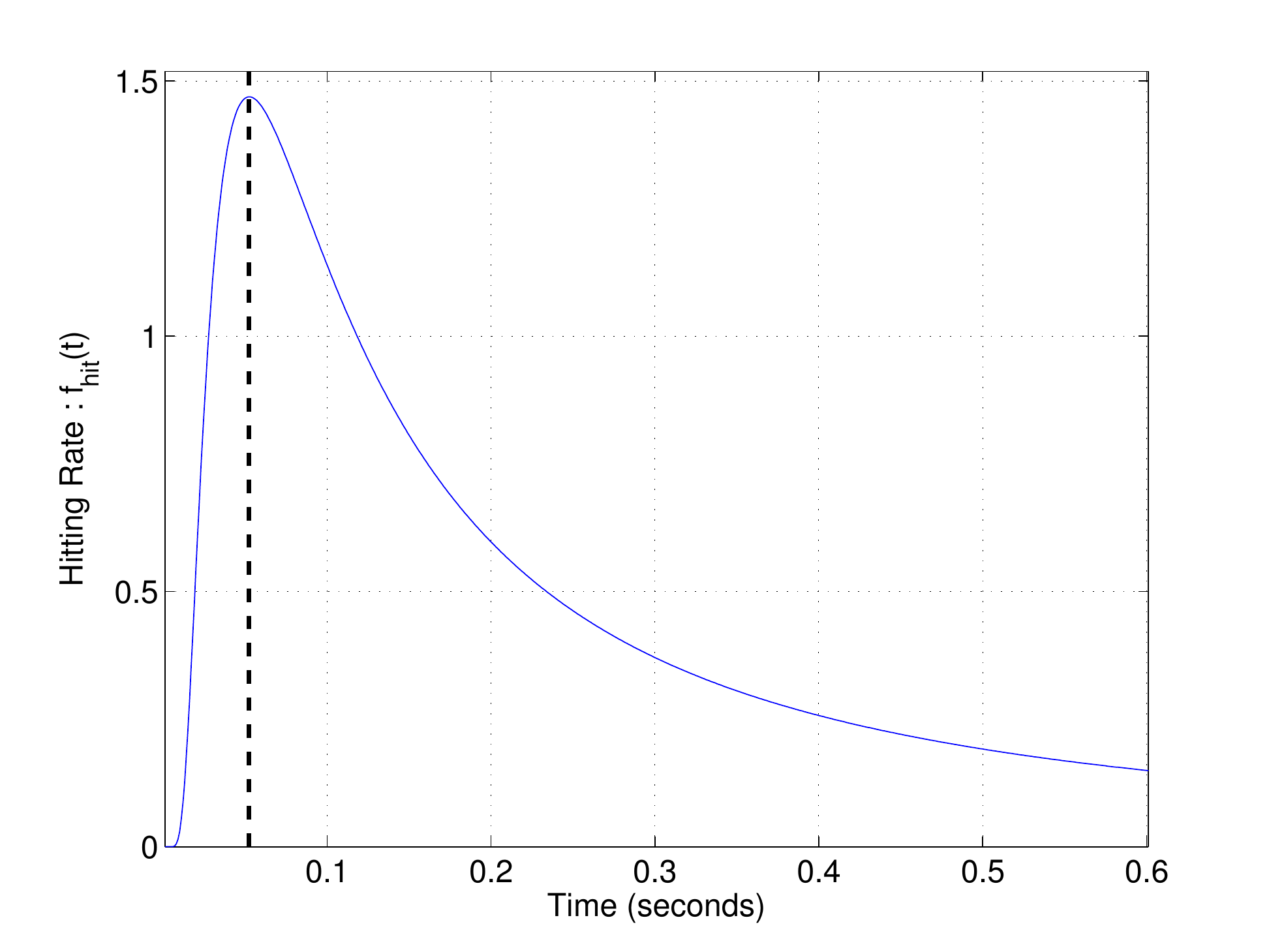}
  \caption{$\fhit$ for $r_r = 5\mu m$, $r_0 = 10\mu m$, and $D= 79.4 \mu m^{2}/s$.}
  \label{fig:fHit}
\end{figure}

Using (\ref{eq:di}), the hitting rate of molecules is also calculated in ~\cite{yilmaz20143dChannelCF} as
\begin{align}
\begin{split}
\label{eqn:fhit}
\fhit &= 4 \pi \rrn^2 \,\, w \,\, p(\rrn, t|r_0) = \displaystyle\frac{\rrn}{r_0} \frac{1}{\sqrt{4 \pi D t}} \frac{r_0 - \rrn}{t} \,\,e^{- \frac{(r_0 - \rrn)^2}{4Dt} },
\end{split}
\end{align}
\noindent which is illustrated in Figure \ref{fig:fHit} for $r_r = 5\mu m$, $r_0 = 10\mu m$, and $D= 79.4 \mu m^{2}/s$ \footnote{These values are considered to be typical, since they simulate an environment such that human insulin hormone is used as the messenger molecules, and a device whose capabilities are similar to a pancreatic $\beta$-cell is used as the tranmistter~\cite{kuran2010energyMF}.}. Notice that $\fhit$ has one peak around $52ms$, where the fraction of absorbed molecules reaches its maximum value. Hence, we can find the mean pulse peak time, $t_{\text{peak}}$, by finding the vanishing point for the derivative of $\fhit$ with respect to time, which leads to
\begin{align}
E[t_{\text{peak}}] = \frac{d^2}{6D}.
\end{align}

\noindent Furthermore, integrating $\fhit$, fraction of molecules absorbed by the receiver until time $t$, $\Fhit$, can be obtained as
\begin{align}
\begin{split}
\label{eqn:Fhit}
\Fhit  = \int\limits_{0}^{t} \fhitt{t'}  dt' = \frac{\rrn}{r_0} \,\, \erfc \left[\frac{r_0 - \rrn}{\sqrt{4Dt}}\right].
\end{split}
\end{align}
Time dependent formulation for the fraction of molecules absorbed by the receiver is an important formulation in the nanonetworking domain, since (\ref{eqn:fhit}) and (\ref{eqn:Fhit}) describe the response of the diffusion channel completely. Additionally, $t_{\text{peak}}$ has a great significance for the choice of symbol duration $t_{s}$ for a nano communication system. In terms of ISI mitigation, it is desirable to have the first hitting probability $p_{1}$ to be the largest in magnitude compared to $p_{k}$ for $k>1$. Therefore, $t_{s}$ should be chosen such that hitting probabilities are in a descending order (\,$p_{1}>p_{2}>p_{3}>...$).

\subsection{Modulation and demodulation techniques} \label{subsec:mod_demod}

BCSK and BMoSK are the two most common modulation techniques for MCvD. In BCSK, number of the received messenger molecules is used as the amplitude of the signal. The receiver decodes the intended symbol as a \bt{1} if the number of messenger molecules arriving at the receiver during a time slot exceeds a pre-determined threshold, and as a \bt{0}, otherwise. To represent different values of symbols, the transmitter releases different number of molecules for each value the symbol can represent, e.g., the transmitter releases $n_0$ molecules for a \bt{0}, whereas it releases $n_1$ molecules for a \bt{1}~\cite{5962989}. As mentioned earlier, in the literature, the threshold is typically empirically chosen by using a long sequence of pilot symbols.     

The BMoSK, on the other hand, utilizes the emission of two different types of messenger molecules to represent information. The transmitter releases a constant number of type-$A$ or type-$B$ molecules for the current symbol values of \bt{0} and \bt{1}, respectively. The receiver then decodes the intended symbol based on the type and the number of the molecules received during a time slot~\cite{5962989}. Unlike CSK, decoding of a MoSK-encoded binary sequence does not necessarily require a threshold value. The receiver can make a decision simply by comparing the received number of molecules of both molecule types and determining which one is larger.

\begin{figure}
\centering
\begin{subfigure}{.5\textwidth}
  \centering
  \includegraphics[width=1\linewidth]{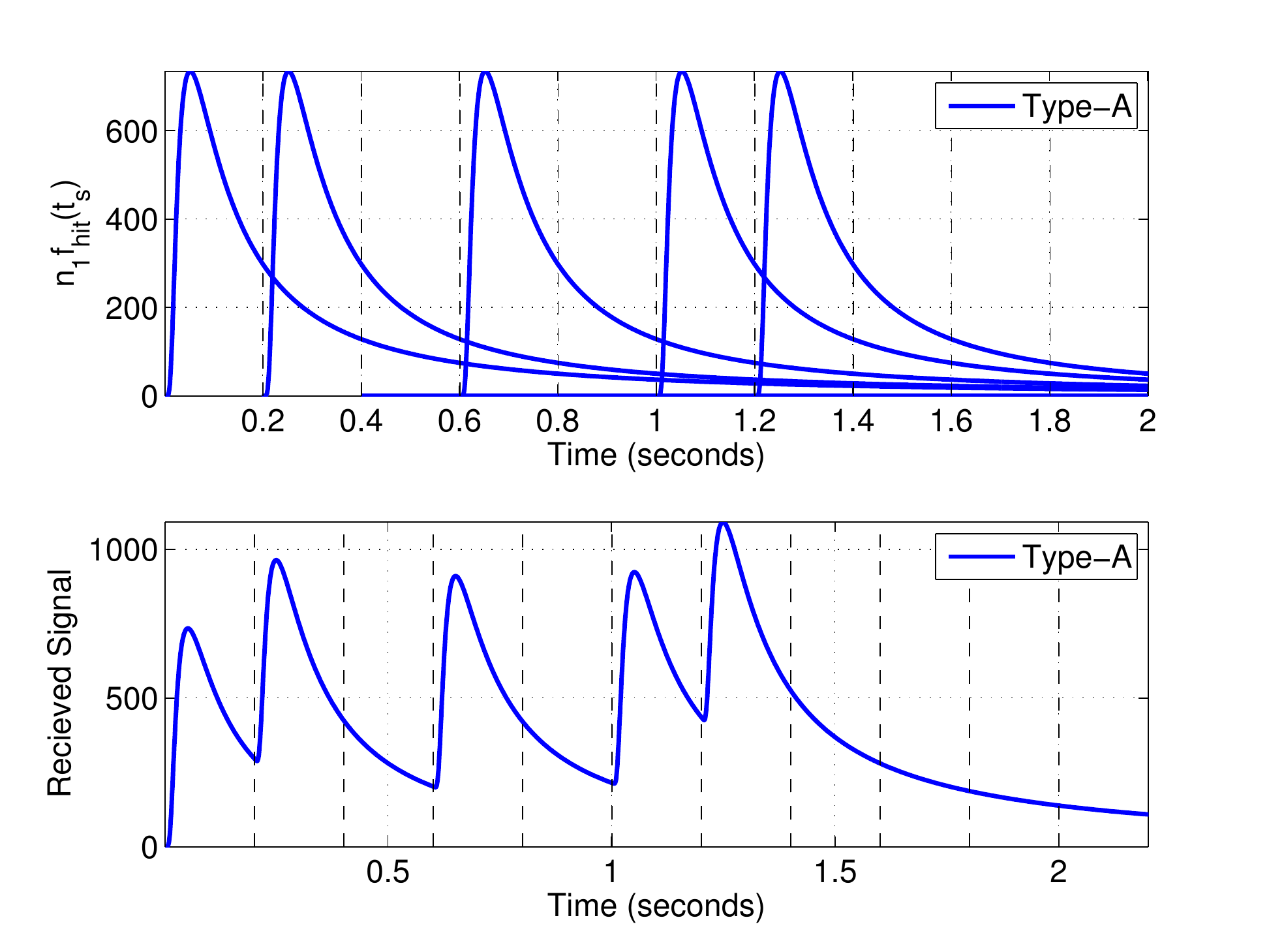}
  \caption{Effects of ISI for BCSK encoded sequence, where $r_r = 5\mu m$, $r_0 = 10\mu m$, $D= 79.4 \mu m^{2}/s$, $n_{0}=0$, $n_{1}=500$ and $t_{s} = 200ms$.}
  \label{fig:ISI_CSK}
\end{subfigure}%
\begin{subfigure}{.5\textwidth}
  \centering
  \includegraphics[width=1\linewidth]{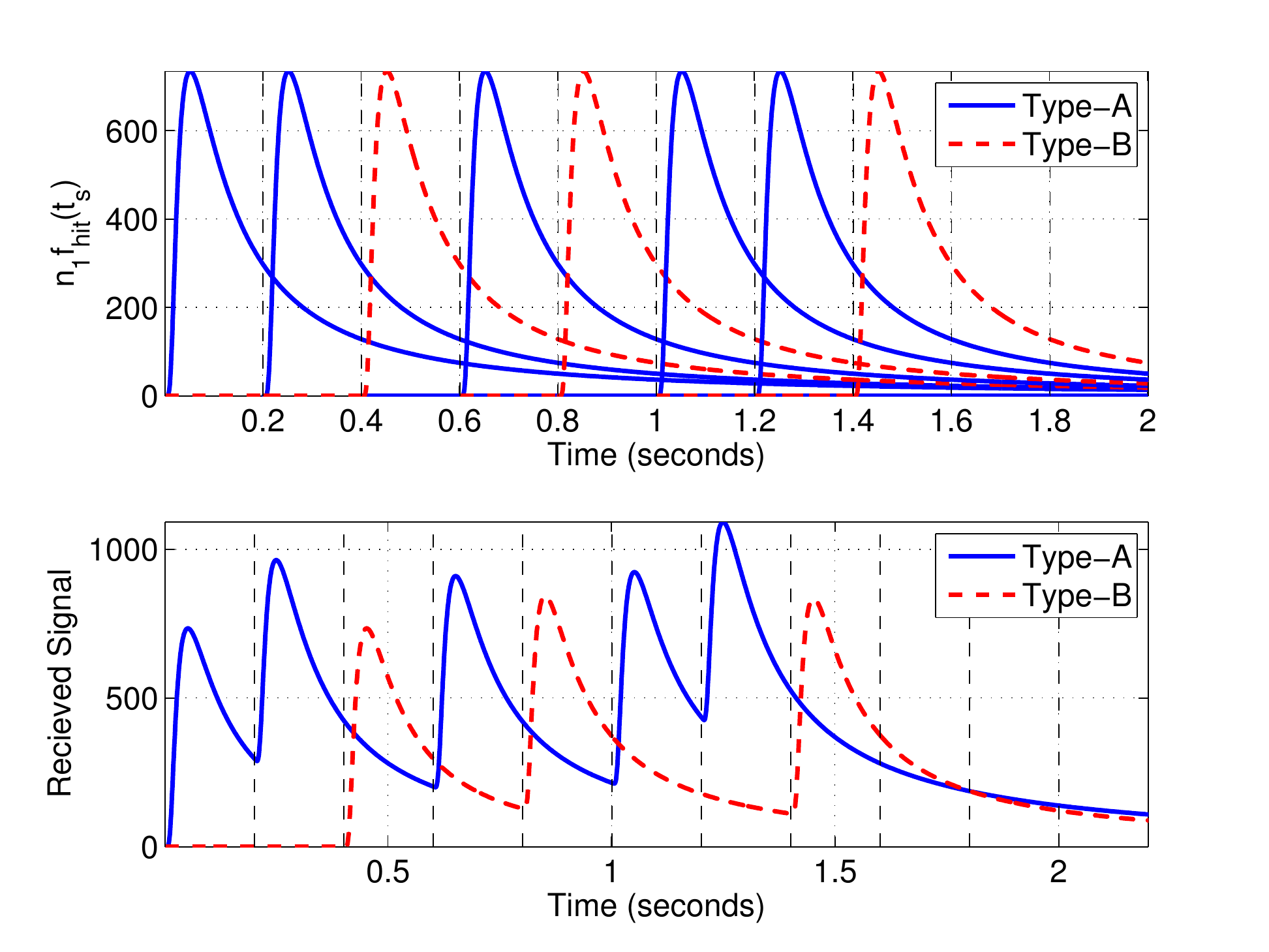}
  \caption{Effects of ISI for BMoSK encoded sequence, where $r_r = 5\mu m$, $r_0 = 10\mu m$, $D= 79.4 \mu m^{2}/s$, $n_{0}=0$, $n_{1}=500$ and $t_{s} = 200ms$.}
  \label{fig:ISI_MOSK}
\end{subfigure}
\caption{Effects of ISI for CSK and MoSK encoded sequences of ${\{1,1,0,1,0,1,1\}}$.}
\label{fig:test}
\end{figure}

The MCvD system using BCSK can be adversely affected from ISI, caused by the residual molecules from the previous symbols~\cite{5962989}. By using (\ref{eqn:fhit}), the hitting rates for a BCSK-encoded binary message sequence of ${\{1,1,0,1,0,1,1\}}$ are calculated, and effects of ISI on each time slot are illustrated in Figure~\ref{fig:ISI_CSK}.

Similar to the BCSK, the residual molecules from the previous symbols also cause ISI when BMoSK is used. BMoSK is less susceptible to ISI effects than the BCSK technique~\cite{5962989}. However, BMoSK requires the synthesis of two types of molecules rather than one, and number of molecules released from transmitter almost doubles, since \bt{0}s are also encoded with constant number of molecules. Effects of ISI for BMoSK is illustrated in Figure \ref{fig:ISI_MOSK}, where the same binary message sequence of ${\{1,1,0,1,0,1,1\}}$ is used.

In conclusion, both modulation techniques are inefficient in terms of energy efficiency and ISI mitigation when molecular communication at high data rates is considered. Additionally, there is no technique in the literature besides the empirical one to calculate the threshold value for a BCSK encoded sequence. 

\section{ISI threshold computation technique}

As stated in the introduction, the optimal threshold is traditionally calculated empirically in the literature. The empirical calculation, however, requires a large sample set of received molecule counts to have a good representation of the system behavior, which means that one must use a long sequence of pilot symbols. In this approach, a slight change in the system model parameters such as temperature, diffusion coefficient, transmitter - receiver distance, etc., requires all these computations to be repeated. This is the main motivation for developing an analytical approach to calculate the optimal threshold.

By using $\Fhit$ given in (\ref{eqn:Fhit}), probability of a single molecule to hit the receiver in a given time slot can be calculated. Let $p_{k}$ denote the hitting probabilities, where $p_1$ is the hitting probability in the current symbol duration and $p_{k}$ for $k \geq 2$ denote the hitting probabilities in the consecutive symbol durations. Hitting probabilities $p_{k}$ for $k=1,2,...$ for a given system model can be calculated using 
\begin{align}
    p_{k} = 
\begin{cases}
    \Fhitt{kt_{s}}-\Fhitt{[k-1]t{s}},& \text{if } k > 1,\\
    \Fhitt{t_{s}},&  \text{if } k = 1,
\end{cases}
\label{eq:hit}
\end{align}

\noindent where $t_ s$ denotes the symbol duration. 

Hitting probabilities are sufficient to describe the characteristics of the diffusion channel completely, which implies that the choice of symbol duration has a great significance in determination of the channel response. In terms of ISI mitigation, it is desirable to have the first hitting probability $p_{1}$ to be the largest in magnitude compared to $p_{k}$ for $k>1$. Therefore, $t_{s}$ should be chosen such that hitting probabilities are in a descending order (\,$p_{1}>p_{2}>p_{3}>...$).

Let $\mathbf{b}_{1}^{n} = \{b_{1},b_{2},...,b_{n}\}$ denote the binary message sequence of length $n$, and let $b_{i}$ and $M_{i}$ denote the message symbols and number of molecules sent from the transmitter in the $i^{th}$ time slot for $i=1,2,...,n$, respectively. For simplicity, assume BCSK, where the number of molecules to be transmitted is $M_{i}=M$ for $b_{i} = 1$, and $M_{i}=0$ for $b_{i} = 0$. 

The number of molecules induced at the receiver for a given time slot can be modeled as a Gaussian random variable~\cite{b01}. Let $\C$ denote the number of molecules induced at the receiver in the $i^{th}$ time slot due to the transmission of $\mathbf{b}_{1}^{i}$. The probability model for $\C$ can be defined~\cite{b01} as
\begin{align}
b_{i} & \sim \mathcal{BE}(P[b_{i}=1]), \label{eq:bern} \\
\C|\mathbf{b}_{1}^{i} &\sim \gauss, \label{eq:gaus}
\end{align}

\noindent where $\mathcal{BE}(\cdot)$, $\mathcal{N}(\cdot,\cdot)$, $P[b_{i}=1]$, and $P[b_{i}=0]$ denote the Bernoulli distribution, Gaussian distribution, probability of occurrence for \bt{1} and probability of occurrence for \bt{0} in the message sequence, respectively. Due to ISI, the expected number of molecules arriving at the receiver in the $i^{th}$ time slot can be given as
\begin{align}
\label{eq:mean01}
\E[\C|\mathbf{b}_{1}^{i}] &= \mu^{\{i\}} = M\sum_{k=1}^{i}p_{k}\,b_{i-k+1},
\end{align}

\noindent which is the mean of the Gaussian distributed molecule count at the receiver. The variance of $\C$ is similarly given by
\begin{align}
\label{eq:std01}
\Var[\C|\mathbf{b}_{1}^{i}] = \sigma^{2\{i\}} = M\sum_{k=1}^{i} p_{k}\,(1-p_{k})\,b_{i-k+1}.
\end{align}

It should be noted that  (\ref{eq:mean01}) and (\ref{eq:std01}) do not include any randomness except for the one due to the diffusion process. This model can be extended by the addition of a zero-mean white Gaussian noise with a constant variance $\sigma^2_{c}$, which may represent the counting noise at the receiver, or noise due to the molecule reactions in the environment, etc. Since the sum of two independent Gaussian random variables is again a Gaussian random variable, with its mean being the sum of the two means, and its variance being the sum of the two variances, the variance of $\C$ becomes
\begin{align}
\label{eq:std02}
\sigma^{2\{i\}} = \sigma_{c}^2 + M\sum_{k=1}^{i}p_{k}\,(1-p_{k})\,b_{i-k+1}.
\end{align}

\begin{figure}
\centering{ \scriptsize
 \begin{tikzpicture}[level distance=0.8cm,
  level 1/.style={sibling distance=8.4cm},
  level 2/.style={sibling distance=4.3cm},
  level 3/.style={sibling distance=2cm}]
  \node {$\cdot$}
     child {node {$\mathbf{d}_{0}^{\{1,0\}} = 0$}
     	 child {node {$\mathbf{d}_{0}^{\{2,0\}}= 00$}
		child {node {$\mathbf{d}_{0}^{\{3,0\}} = 000$}} 
      		child {node {$\mathbf{d}_{1}^{\{3,0\}} = 001$}}}
      	child {node {$\mathbf{d}_{1}^{\{2, 0\}} = 01$} 
		child {node {$\mathbf{d}_{0}^{\{3,2\}} = 010$}} 
      		child {node {$\mathbf{d}_{1}^{\{3,2\}} = 011$}}}
      }
     child {node {$\mathbf{d}_{1}^{\{1,0\}} = 1$}
     	 child {node {$\mathbf{d}_{0}^{\{2,1\}}= 10$}
		child {node {$\mathbf{d}_{0}^{\{3,1\}} = 100$}} 
      		child {node {$\mathbf{d}_{1}^{\{3,1\}} = 101$}}}
      	child {node {$\mathbf{d}_{1}^{\{2,1\}} = 11$} 
		child {node {$\mathbf{d}_{0}^{\{3,3\}} = 110$}} 
      		child {node {$\mathbf{d}_{1}^{\{3,3\}} = 111$}}}
      };
\end{tikzpicture}}
\caption{Binary tree for $n=3$.} \label{fig:binTree}
\end{figure}
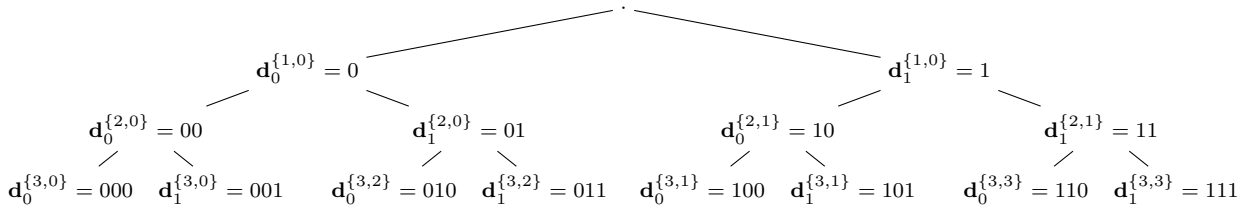

Equations (\ref{eq:mean01}) and (\ref{eq:std01}) indicate that the parameters of the Gaussian distribution change for each symbol, which means that each and every symbol requires its own optimal threshold. To begin with, let us focus on finding the optimal threshold $\gamma^{*\{i\}}$ for the detection of ${b}_{i}$ in $\mathbf{b}_{1}^{n}$ such that

\begin{align}
\label{eq:dec}
\C \underset{\hat{b}_{i} = 0}{\overset{\hat{b}_{i} = 1}{\gtrless}} \gamma^{*\{i\}},
\end{align}

\noindent where $\hat{b}_{i}$ denotes the estimate of $b_{i}$. We can treat this case as a traditional binary detection problem in an AWGN channel and use \emph{maximum a posteriori probability} (MAP) decision rule, given as 
\begin{align}
\label{eq:MAP01}
\frac{P[b_{i}=1]}{P[b_{i}=0]} \frac{p(C_{i}|\mathbf{b}_{1}^{i-1},b_{i}=1)}{p(C_{i}| \mathbf{b}_{1}^{i-1},b_{i}=0)} \underset{\hat{b}_{i} = 0}{\overset{\hat{b}_{i} = 1}{\gtrless}} 1,
\end{align}
\noindent where $p(\cdot)$ denotes the probability density function of the Gaussian distributed $\C$. Note that calculation of $\gamma^{*\{i\}}$ requires information about the sequence history $\mathbf{b}_{1}^{i-1}$. In case of a memoryless decoder, when the first $(i-1)$ bits, which are crucial for (\ref{eq:mean01}) and (\ref{eq:std01}), are unknown, all possible combinations (candidates) for $\mathbf{b}_{1}^{i-1}$ must be considered, which yields to $2^{i-1}$ candidate means, variances, and optimal thresholds. Each $\mathbf{b}_{1}^{i-1}$ candidate also has its own probability, $P[\mathbf{b}_{1}^{i-1}]$. To allow for the enumeration of the candidates, each candidate sequence is denoted by $\mathbf{d}^{\{i, j\}}_{k}$, and its corresponding optimal threshold is denoted by ${\gamma^{*\{i,j\}}}$. In this notation, $j$ is equal to the decimal value of reverse ordered $\mathbf{b}_{1}^{i-1}$ sequence, $i$ denotes the length of the candidate sequence, and $k$ denotes whether the candidate sequence is conditioned on $b_{i} = 0$ or $b_{i}=1$. For example, $\mathbf{d}^{\{3, 1\}}_{0}$ corresponds to the bit sequence of length three, conditioned on \bt{0}, and with decimal value $1$ for the reverse ordered history; which can only be the sequence $\{1\,0\,0\}$. Reversing the order of $\mathbf{b}_{1}^{i-1}$ becomes significant when we consider the effects of ISI, since the latter symbols contribute more to the ISI than the former ones. Thus, $j$ allows for the ordering of the amount of ISI for each  $\mathbf{d}^{\{i, j\}}_{k}$. Possible candidate sequences for $n=3$ can be visualized as a binary tree given in Figure \ref{fig:binTree}. 

By conditioning each candidate on \bt{1} and \bt{0}, the optimal threshold $\gamma^{*\{i,j\}}$ can be found for siblings $\{\mathbf{d}_{0}^{\{i,j\}},\mathbf{d}_{1}^{\{i,j\}}\}$ in the binary tree, which are basically two Gaussian distributions with parameter sets $\{\mu_{0}^{\{i,j\}}, \sigma_{0}^{\{i,j\}}\}$ and $\{\mu_{1}^{\{i,j\}}, \sigma_{1}^{\{i,j\}}\}$, respectively. Each $\gamma^{\{i,j\}}$ can be calculated by writing (\ref{eq:MAP01}) and (\ref{eq:dec}) explicitly, resulting in the quadratic equation

\begin{align}
a\left(\gamma^{*\{i\}}\right)^{2} +  b\gamma^{*\{i\}} + c = 0,
\label{eq:quad}
\end{align}
\noindent where,
\begin{align}
a &=  \left[{\sigma_{1}^{2\{i,j\}} - \sigma_{0}^{2\{i,j\}}}\right], \\
b &= 2\left[{\sigma_{0}^{2\{i,j\}}\mu_{1}^{\{i,j\}}- \sigma_{1}^{2\{i,j\}}}\mu_{0}^{\{i,j\}}\right], \\
c &= \sigma_{1}^{2\{i,j\}}\mu_{0}^{2\{i,j\}} - \sigma_{0}^{2\{i,j\}}\mu_{1}^{2\{i,j\}} - 2\sigma_{1}^{2\{i,j\}}\sigma_{0}^{2\{i,j\}}\log\left[\frac{P[b_{i}=0]\sigma_{1}^{\{i,j\}}}{P[b_{i}=1]\sigma_{0}^{\{i,j\}}}\right].
\end{align}
\noindent This equation can be solved analytically considering the positive root $\gamma^{*\{i\}} = \frac{-b+\sqrt{\Delta}}{2a}$, where $\Delta = b^2 - 4ac$. 

To find the optimal threshold $\gamma^{*\{i\}}$ that minimizes the overall probability of error in the detection of ${b}_{i}$ considering all candidate sequences, the minimizing function can be written as
\begin{align}
\label{eq:sum01}
J_{i}(\gamma) = \sum_{j=1}^{2^{i-1}-1} P[\mathbf{d}_{0}^{\{i,j\}}]Q\left({\frac{\gamma - \mu_{0}^{\{i,j\}}}{\sigma_{0}^{\{i,j\}}}}\right) +  P[\mathbf{d}_{1}^{\{i,j\}}]Q\left({\frac{\mu_{1}^{\{i,j\}}-\gamma}{\sigma_{1}^{\{i,j\}}}}\right), \,\, \forall i\geq 1,
\end{align}
which is equal to the sum of error probabilities for a given threshold $\gamma$. To minimize $J_{i}(\gamma)$, derivative with respect to $\gamma$ can be set to zero as
\begin{align} 
\frac{\partial J_{i}(\gamma)}{\partial \gamma} &= \sum_{j=1}^{2^{i-1}-1} P[\mathbf{d}_{1}^{\{i,j\}}]\mathcal{N}(\gamma^{*\{i\}} | \mu_{1}^{\{i,j\}}, \sigma_{1}^{2\{i,j\}}) - P[\mathbf{d}_{0}^{\{i,j\}}]\mathcal{N}(\gamma^{*\{i\}} | \mu_{0}^{\{i,j\}}, \sigma_{0}^{2\{i,j\}}) = 0,
\end{align}
which yields 
\begin{align}
\label{eq:sums01}
 \sum_{j=1}^{2^{i-1}-1} P[\mathbf{d}_{1}^{\{i,j\}}]\mathcal{N}(\gamma^{*\{i\}} | \mu_{1}^{\{i,j\}}, \sigma_{1}^{2\{i,j\}}) =  \sum_{i=1}^{2^{i-1}-1} P[\mathbf{d}_{0}^{\{i,j\}}]\mathcal{N}(\gamma^{*\{i\}} | \mu_{0}^{\{i,j\}}, \sigma_{0}^{2\{i,j\}}).
\end{align}
For $i>2$, (\ref{eq:sums01}) becomes a hard problem to solve analytically, so numerical methods are used instead. 

$\gamma^{*\{{i}\}}$ can be efficiently computed using two fundamental observations. The first observation makes use of (\ref{eq:mean01}), (\ref{eq:std01}), and the fact that $\ts$ is chosen such that hitting probabilities are in an descending order, i.e., $p_{i}>p_{j}$ for $i<j$. We can then sort the distribution parameters as
\begin{align}
\mu_{0}^{\{i,1\}} < \mu_{0}^{\{i,2\}}& < ... < \mu_{0}^{\{i,2^{i-1}-1\}}, \nonumber \\
\mu_{1}^{\{i,1\}} < \mu_{1}^{\{i,2\}}&< ... < \mu_{1}^{\{i,2^{i-1}-1\}}, \nonumber \\
\sigma_{0}^{\{i,1\}} < \sigma_{0}^{\{i,2\}}&< ... < \sigma_{0}^{\{i,2^{i-1}-1\}}, \nonumber \\
\sigma_{1}^{\{i,1\}} < \sigma_{1}^{\{i,2\}}&< ... < \sigma_{1}^{\{i,2^{i-1}-1\}}.
\end{align}
Consequently, optimal thresholds can also be sorted as 
\begin{align}
\gamma^{*\{i,1\}} < \gamma^{*\{i,2\}} < ... < \gamma^{*\{i,2^{i-1}-1\}}.
\end{align}
Being able to sort these optimal thresholds provides us with an upper bound  for $\gamma^{*\{i,2^{i-1}-1\}}$, since the optimal threshold considering all candidates cannot be greater than the optimal threshold considering only the siblings $\{\mathbf{d}_{0}^{\{i,2^{i-1}-1\}},\mathbf{d}_{1}^{\{i,2^{i-1}-1\}}\}$ with the highest mean and variance values, i.e., $\gamma^{*\{i\}}< \gamma^{*\{i,2^{i-1}-1\}}$. 

The second observation is that, as $n$ increases, due to the accumulating molecules in the diffusion channel, the optimal threshold also increases monotonically, i.e., $\gamma^{*\{i-1\}}<\gamma^{*\{i\}}$. In conclusion, the lower and upper bounds for $\gamma^{*\{i\}}$ can be written as
\begin{align}
\gamma^{*\{i-1\}}<\gamma^{*\{i\}} < \gamma^{*\{2^{i-1}-1\}}.
\end{align}

\noindent These bounds allow for the search of $\gamma^{*\{i\}}$ by employing a fixed point iteration. Such an iterative algorithm is given in Algorithm \ref{alg:th}.
\begin{algorithm}[t]
\begin{algorithmic}[1]
\STATE Compute $\gamma^{*\{1\}}$ 
\FOR{$i=2$ to $N$}
\FOR{$j=0$ to $2^{i-1}-1$ }
\STATE Calculate $\{\mu_{0}^{\{i,j\}},\mu_{1}^{\{i,j\}},\sigma_{0}^{\{i,j\}},\sigma_{1}^{\{i,j\}}\}$
\IF{$j=2^{i-1}-1$} 
\STATE Compute optimal threshold $\gamma^{*\{i,2^{i-1}-1\}}$
\STATE $\gamma_{max} \leftarrow \gamma^{*\{i,2^{i-1}-1\}}$ 
\ENDIF
\ENDFOR
\STATE Set step size $\alpha = 0.1$
\STATE $\boldsymbol{\gamma} = \gamma^{*\{i-1\}}:\alpha:\gamma_{max}$ 
\WHILE{$\alpha > 10^{-4}$} 
\STATE{Calculate the sum of likelihoods \\[1.2ex]
$ \displaystyle \mathbf{r} = \frac{ \sum_{j=1}^{2^{i-1}-1}P[b_{i}=1]\mathcal{N}(\boldsymbol\gamma | \mu_{1}^{\{i,j\}}, \sigma_{1}^{\{i,j\}})}{\sum_{j=1}^{2^{i-1}-1}P[b_{i}=0]\mathcal{N}(\boldsymbol\gamma | \mu_{0}^{\{i,j\}}, \sigma_{0}^{\{i,j\}})}$ \\[1.2ex]
}
\STATE{$ \displaystyle m^{*} \leftarrow \argmin_m |1-r(m)|$ \\[1ex]}
\STATE{$\gamma^{*\{i\}} \leftarrow \boldsymbol{\gamma}(m^{*})  $}
\STATE{$\boldsymbol{\gamma} \leftarrow \gamma^{*\{i\}}-\alpha:\alpha / 10 : \gamma^{*\{i\}}+\alpha$}
\STATE{$\alpha \leftarrow \alpha / 10$}
\ENDWHILE
\ENDFOR
\end{algorithmic}
\caption{Calculation of $\gamma^{*\{N\}}$}
\label{alg:th}
\end{algorithm}

\begin{example}
Consider a BCSK modulated random binary sequence, where $P[b_{i} = 0] = P[b_{i} = 1] = 0.5$, $\forall i$. To calculate the optimal threshold $\gamma^{*\{2\}}$, the distribution parameters must be calculated by using (\ref{eq:mean01}) and (\ref{eq:std01}). Candidate sequences and their corresponding parameters are given in Table \ref{table:params}. \\
\begin{table} \centering
\caption{Distribution parameters.}
\begin{tabular}{ccccc} 
Candidate Sequence                             & Mean                               & Standart Deviation                                &Probability of Occurance    &Optimal Threshold   \\   \hline \\[-1.5ex]
$\mathbf{d}_{0}^{\{2,0\}} = \{0,0\}$      & $\mu_{0}^{\{2,0\}} $      & $\sigma^{\{2,0\}}_{0}$         & $0.25$                           							       & \rdelim\}{2}{12mm}[$\gamma^{*\{2,0\}}$] \phantom{00} \rdelim\}{4}{5mm}[$\gamma^{*\{2\}}$]  \\[0.3ex]
$\mathbf{d}_{1}^{\{2,0\}} = \{0,1\}$      & $\mu_{1}^{\{2,0\}} $      & $\sigma^{\{2,0\}}_{1}$         & $0.25$                             						       \\[0.3ex]
$\mathbf{d}_{0}^{\{2,1\}} = \{1,0\}$      & $\mu_{0}^{\{2,1\}} $      & $\sigma^{\{2,1\}}_{0}$         & $0.25$                           							       & \rdelim\}{2}{18mm}[$\gamma^{*\{2,1\}}$] \phantom{00}   \\[0.3ex]
$\mathbf{d}_{1}^{\{2,1\}} = \{1,1\}$      & $\mu_{1}^{\{2,1\}} $      & $\sigma^{\{2,1\}}_{1}$         & $0.25$                               							\\[0.3ex]
\end{tabular}
\label{table:params}
\end{table} \\

To calculate $\gamma^{*\{2\}}$,  (\ref{eq:mean01}) and (\ref{eq:std01}) must be used, which means that information of the first two hitting probabilities are required. Let us define a parameter set, denoted by \set, where $r_r = 5\mu m$, $r_0 = 10\mu m$, $\sigma^{2}_{c}=1$, molecules similar to insulin hormone are used as information carriers, and the channel is filled with a liquid which results in a diffusion coefficient of $79.4 \mu m^{2}/s$~\cite{kuran2010energyMF}. This parameter set will be used for all the latter simulations and examples in this paper. For this example, in addition to the set of parameters \set, the symbol duration is chosen as $t_s = 200 ms$, and $M = 100$ molecules are used as messenger molecules on each symbol duration . With these parameters, hitting probabilities are calculated as $\{p_{1},p_{2} \} = \{0.1875,0.0777\}$. Using these probabilities, the mean and the variance of four possible candidates can be calculated as

\begin{minipage}[t]{.16\textwidth}
\phantom{0000}
\end{minipage}
\begin{minipage}[t]{.26\textwidth}
\begin{align*}
\{\mu_{0}^{\{2,0\}},\sigma^{2\{2,0\}}_{0} \}& =  \{0, 1\}, \\
\{\mu_{0}^{\{2,1\}},\sigma^{2\{2,1\}}_{0} \}& = \{7.77, 8.17\},
\end{align*}
\end{minipage}
\begin{minipage}[t]{.16\textwidth}
\phantom{0000000000}
\end{minipage}
\begin{minipage}[t]{.26\textwidth}
\begin{align*}
\{\mu_{1}^{\{2,0\}},\sigma^{2\{2,0\}}_{1} \}& = \{8.75,16.23\}, \\
\{\mu_{1}^{\{2,1\}},\sigma^{2\{2,1\}}_{1} \}& = \{26.52,23.40\}.
\end{align*}
\end{minipage}
\begin{minipage}[t]{.16\textwidth}
\phantom{0000000000}
\end{minipage}

\phantom{0}

According to (\ref{eq:MAP01}) and (\ref{eq:sums01}), the optimal threshold values must satisfy
\begin{align}
0.5 \mathcal{N}(\gamma^{*\{2,0\}} | \mu_{0}^{\{2,0\}}=0,\sigma^{2\{2,0\}}_{0}=1) &= 0.5 \mathcal{N}(\gamma^{*\{2,0\}} | \mu_{1}^{\{2,0\}}=8.75,\sigma^{2\{2,0\}}_{1}=16.23), \label{eq:n1} \\
0.5 \mathcal{N}(\gamma^{*\{2,1\}} | \mu_{0}^{\{2,1\}}=7.77, \sigma^{2\{2,1\}}_{0}=8.17) &= 0.5 \mathcal{N}(\gamma^{*\{2,1\}} | \mu_{1}^{\{2,1\}}=26.52, \sigma^{2\{2,1\}}_{1}=23.40),\label{eq:n2}
\end{align}
\begin{align}
\begin{split}
0.25 \mathcal{N}(\gamma^{*\{2,0\}} | \mu_{0}^{\{2,0\}}=0,\sigma^{2\{2,0\}}_{0}=1) + 0.25\mathcal{N}&(\gamma^{*\{2,1\}} | \mu_{0}^{\{2,1\}}=7.77, \sigma^{2\{2,1\}}_{0}=8.17)= \\ 0.25  \mathcal{N}(\gamma^{*\{2,0\}} | \mu_{1}^{\{2,0\}}=8.75,\sigma^{2\{2,0\}}_{1}=16.23) &+ 0.25\mathcal{N}(\gamma^{*\{2,1\}} | \mu_{1}^{\{2,1\}}=26.52, \sigma^{2\{2,1\}}_{1}=23.40).\label{eq:n3}
\end{split}
\end{align}

Equations (\ref{eq:n1}) and (\ref{eq:n2}) can be solved analytically, whereas solving for $\gamma^{*\{2\}}$ in (\ref{eq:n3}) requires the use of Algorithm \ref{alg:th}. As a result,
$\gamma^{*\{2,0\}} = 4.0189$, $\gamma^{*\{2,1\}} = 15.1198$, and $\gamma^{*\{2\}} =12.7882$ are obtained.

\end{example}

\begin{figure}
\centering
  \includegraphics[width=.7\columnwidth]{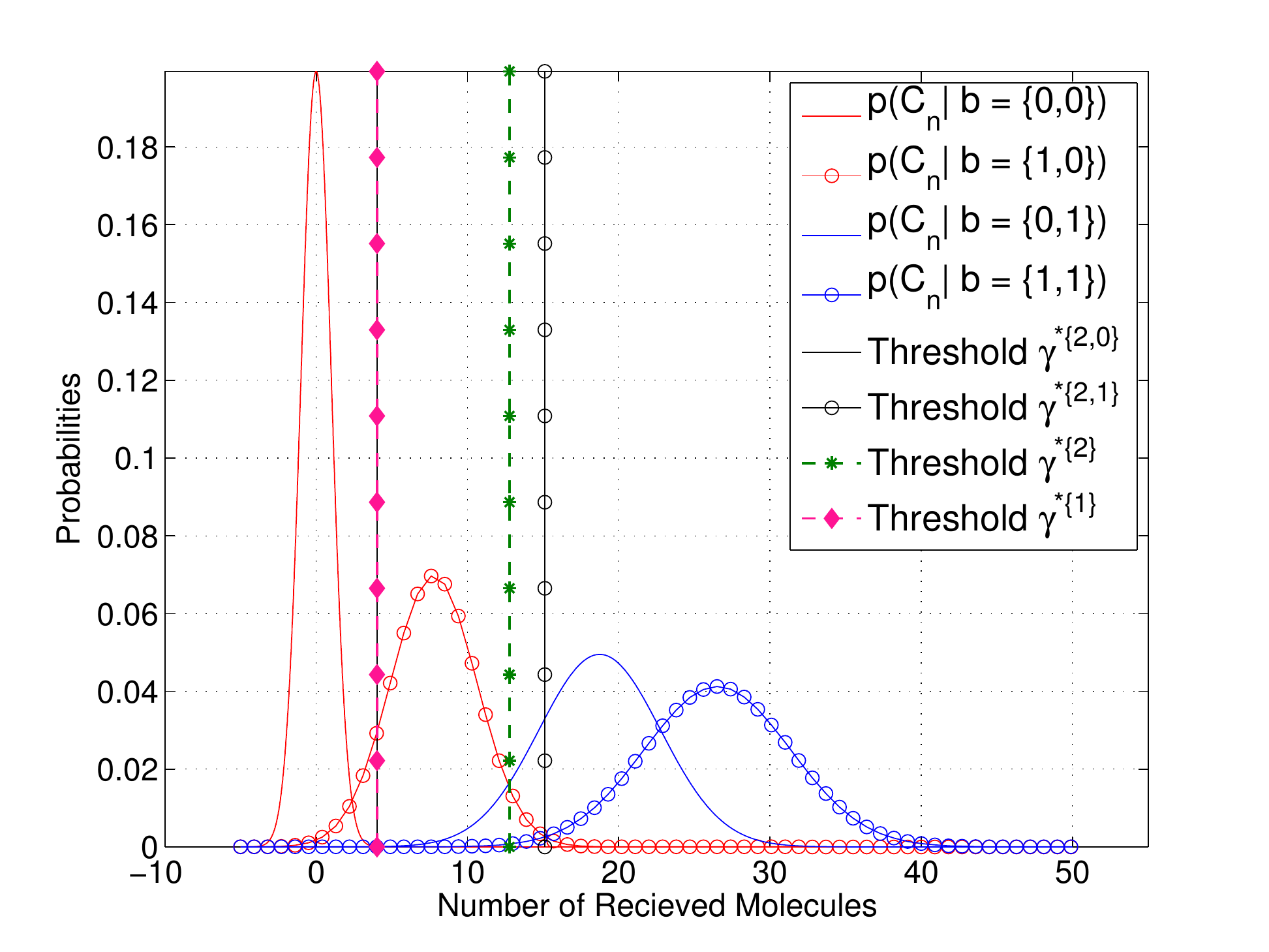}
  \caption{Distributions of recieved molecule counts and threshold values.}
  \label{fig:dist_n_2}
\end{figure}

These results can be interpreted as follows. If the receiver is memoryless, it has to consider all possible candidate sequences in order to make a decision for ${b}_{2}$. In this case, $\gamma^{*\{2\}}$ must be used, meaning that any number of molecules below approximately $13$ molecules yields to a decision of $\hat{b}_{2}=0$. On the other hand, if the receiver stores information about $\hat{b}_{1}$, depending on the binary value of $\hat{b}_{1}$, either $\gamma^{*\{2,0\}}$ or  $\gamma^{*\{2,1\}}$ can be used. Distribution of candidate sequences and threshold values for this example are plotted in Figure \ref{fig:dist_n_2}.


\subsection{Least Mean Squares Regression}
\begin{figure}
\centering
\begin{minipage}[t]{.5\textwidth}
  \centering
  \includegraphics[width=1\columnwidth]{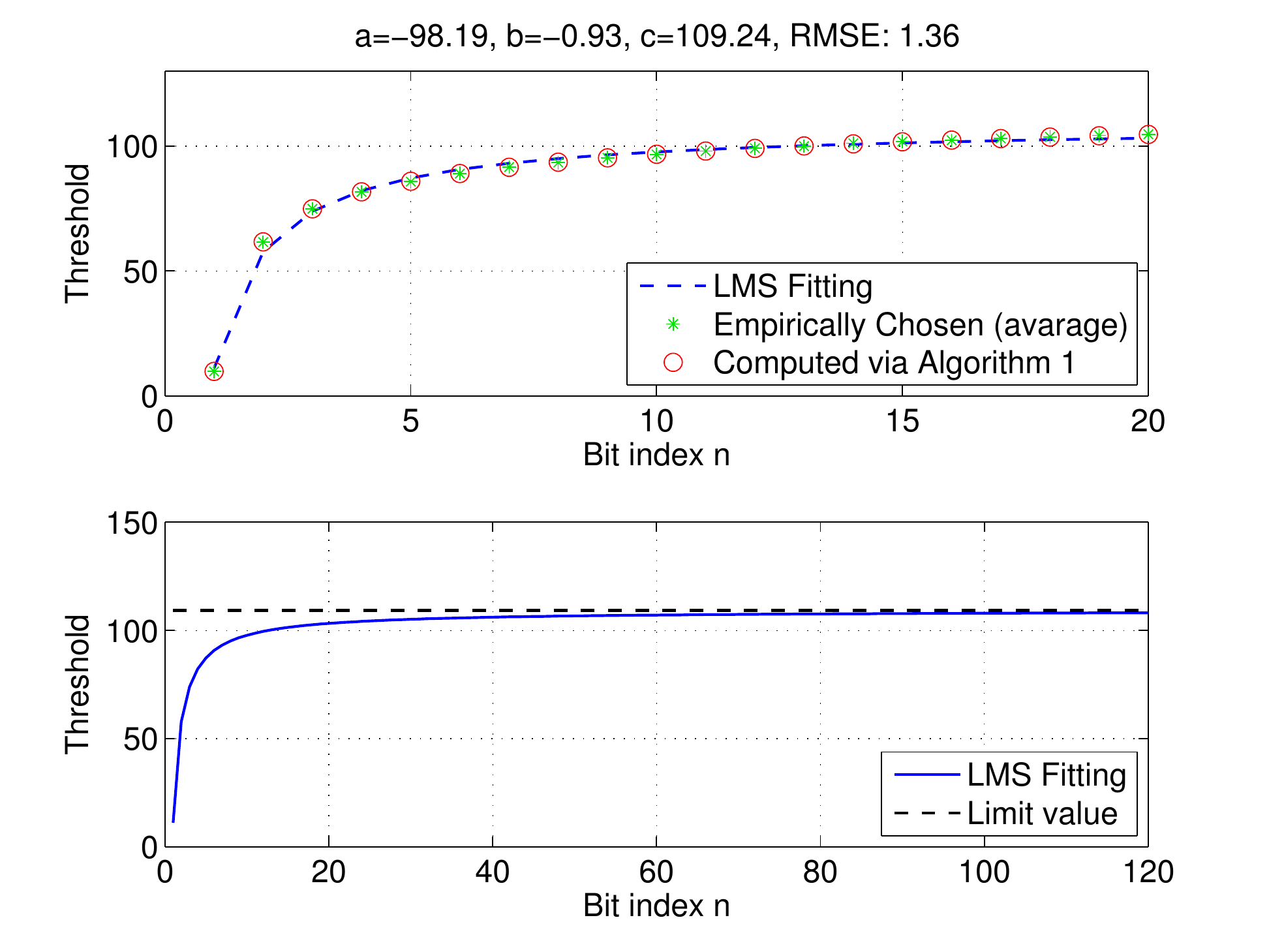}\\
  \caption{Least squares regression for threshold values, where $P[b_{i}=0] = 0.5$, $t_{s} = 200ms$, and $M= 500$.}
  \label{fig:LMS_FIT}
\end{minipage}%
\begin{minipage}[t]{.5\textwidth}
  \centering
  \includegraphics[width=1\columnwidth]{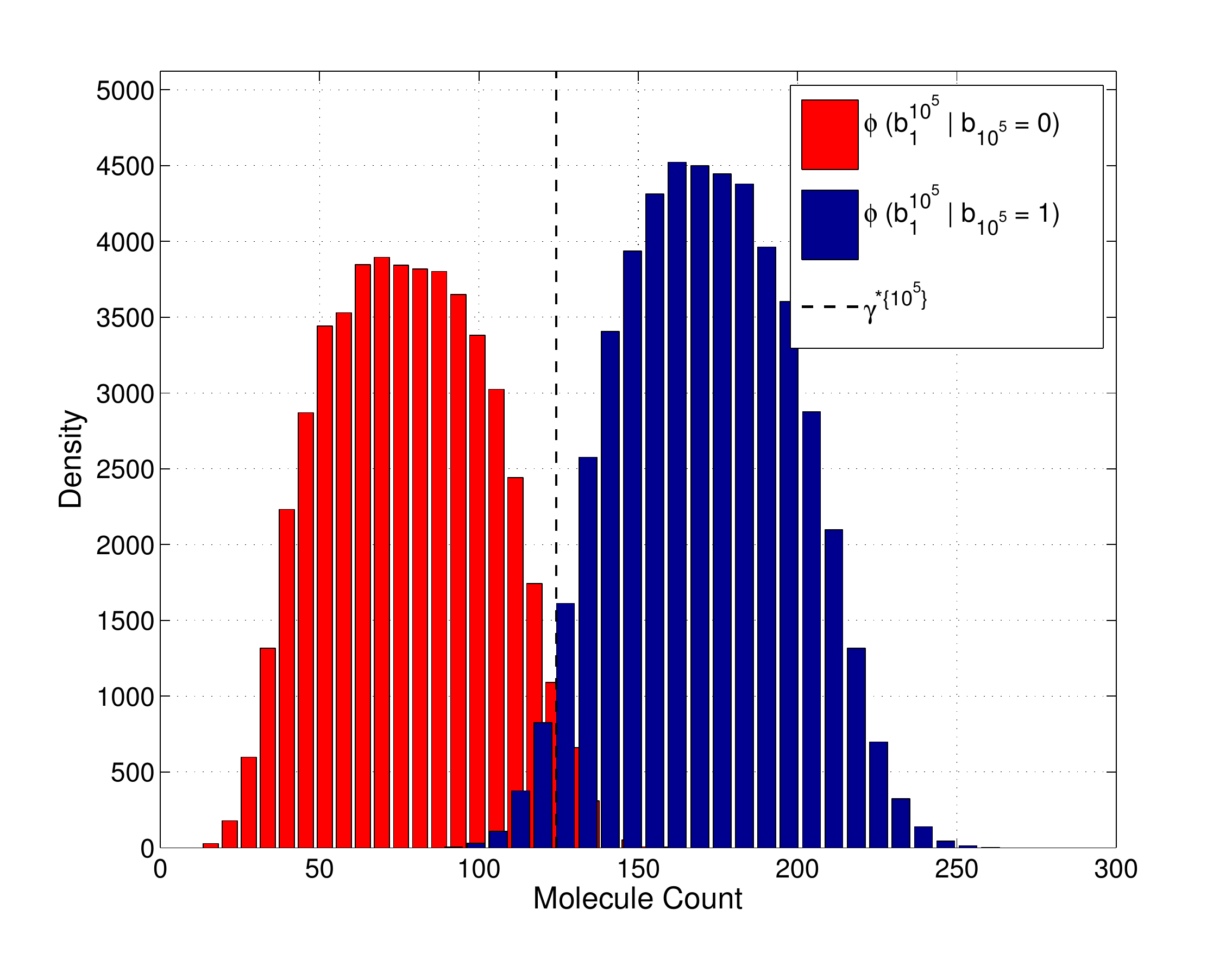}\\
  \caption[]{Histograms of simulated data and optimal threshold value calculated by LMS regression, where $P[b_{i}=0] = 0.5$, $t_{s} = 200ms$, and $M= 500$.}
  \label{fig:TH_HIST}
\end{minipage}
\end{figure}

Computing the optimal threshold for ${b}_{i}$ requires computation of the distribution parameters for all candidates, which means that the number of operations increases with powers of $2$ as $i$ increases. To calculate the optimal thresholds for large values of $i$, least mean squares (LMS) regression can be applied to the threshold values $\gamma^{*\{i\}}$ for $i \leq 20$, using a function of the form
\begin{align}
\gamma^{*}_{i} = \alpha i^{\beta} + \kappa,
\label{eq:lms}
\end{align}
where $\alpha,\beta,\kappa \in \mathbb{R}$ and $-1<\beta<0$. The motivation of using a function of this from can be justified based on two facts. First, note that hitting probabilities have a cumulative effect on both distribution parameters, as seen in (\ref{eq:mean01}) and (\ref{eq:std01}). As $i$ increases, the marginal effect of hitting probabilities will decrease, indicating that there should be a limit value as $i$ goes to infinity. Physical interpretation of this is as follows: since the channel is assumed to be free of molecules before the transmission begins, the optimal threshold value will increase in the early stages of the transmission. As transmission continues, due to the accumulation of the molecules in the diffusion channel, number of molecules in the channel will go into saturation and the optimal threshold value will converge to a constant. The first $20$ threshold values computed using Algorithm \ref{alg:th} and the values for $i>20$ obtained via LMS are shown in Figure \ref{fig:LMS_FIT}. Empirically chosen threshold values for $i \leq 20$ are also included in Figure \ref{fig:LMS_FIT}. Root mean square error (RMSE) is calculated to evaluate the performance of the fit. 

To verify the reliability of the LMS outputs, random binary messages of length $10^5$ consisting of equally likely bits are generated and the histograms of molecule counts conditioned on $b_{10^5} = 0$ and $b_{10^5} = 1$ are plotted. As seen in Figure \ref{fig:TH_HIST}, the optimal threshold $\gamma^{*\{10^5\}}$ computed via LMS regression is at the intersection of two distributions where likelihoods are equal to each other. $\phi(\cdot)$ in Figure \ref{fig:TH_HIST} denotes the unnormalized density of the number of received molecules.

\begin{figure}
\centering
 \includegraphics[width=.6\columnwidth]{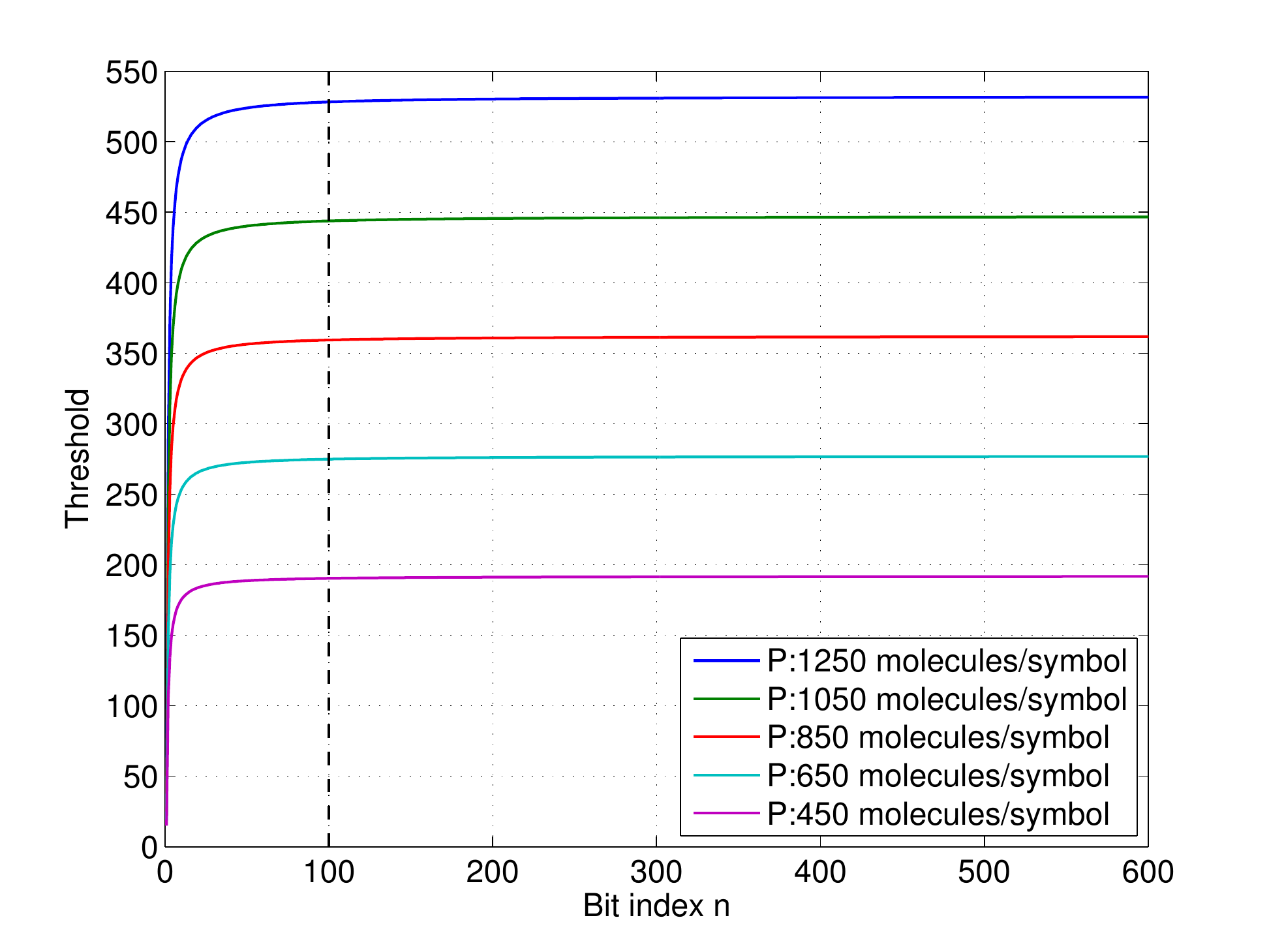}\\
\caption{Optimal threshold values for varying signal power, where $M=100$ and $t_s=200ms$.}
\label{fig:SIG_POW_TH}
\end{figure}

It should be noted that $\alpha$, $\beta$, and $\kappa$ in (\ref{eq:lms}) are dependent both on signal power and hitting probabilities, which are both determined by the environmental parameters, such as diffusion coefficient, receiver radius, etc. Threshold values for different signal power levels are given in Figure \ref{fig:SIG_POW_TH}. 

Since it is possible to compute $\gamma^{*\{i\}}$, $\forall i$, and $\lim_{i \to \inf} \gamma^{*\{i\}} = \kappa$, different strategies can be used for thresholding. As seen in Figure \ref{fig:SIG_POW_TH}, approximately after $100$ bits, all threshold values seem to converge to a constant value, which means that, in case of a continuous transmission, $\kappa$ can be applied as a threshold for $i > 100$. On the other hand, if the decoder cannot afford to store the $\gamma^{*\{i\}}$ for $i \leq 100$, $\mathbf{b}_{1}^{100}$ can be used as a burn-in period. It is also possible to apply different techniques such as binning the bit indices and applying particular threshold values for these particular ranges. Being able to compute $\gamma^{*\{i\}}$, $\forall i$ allows us to apply different thresholding strategies depending on the performance specifications and technical constraints.  

\begin{figure}
\centering
\begin{minipage}[t]{.5\textwidth}
  \centering
  \includegraphics[width=1\columnwidth]{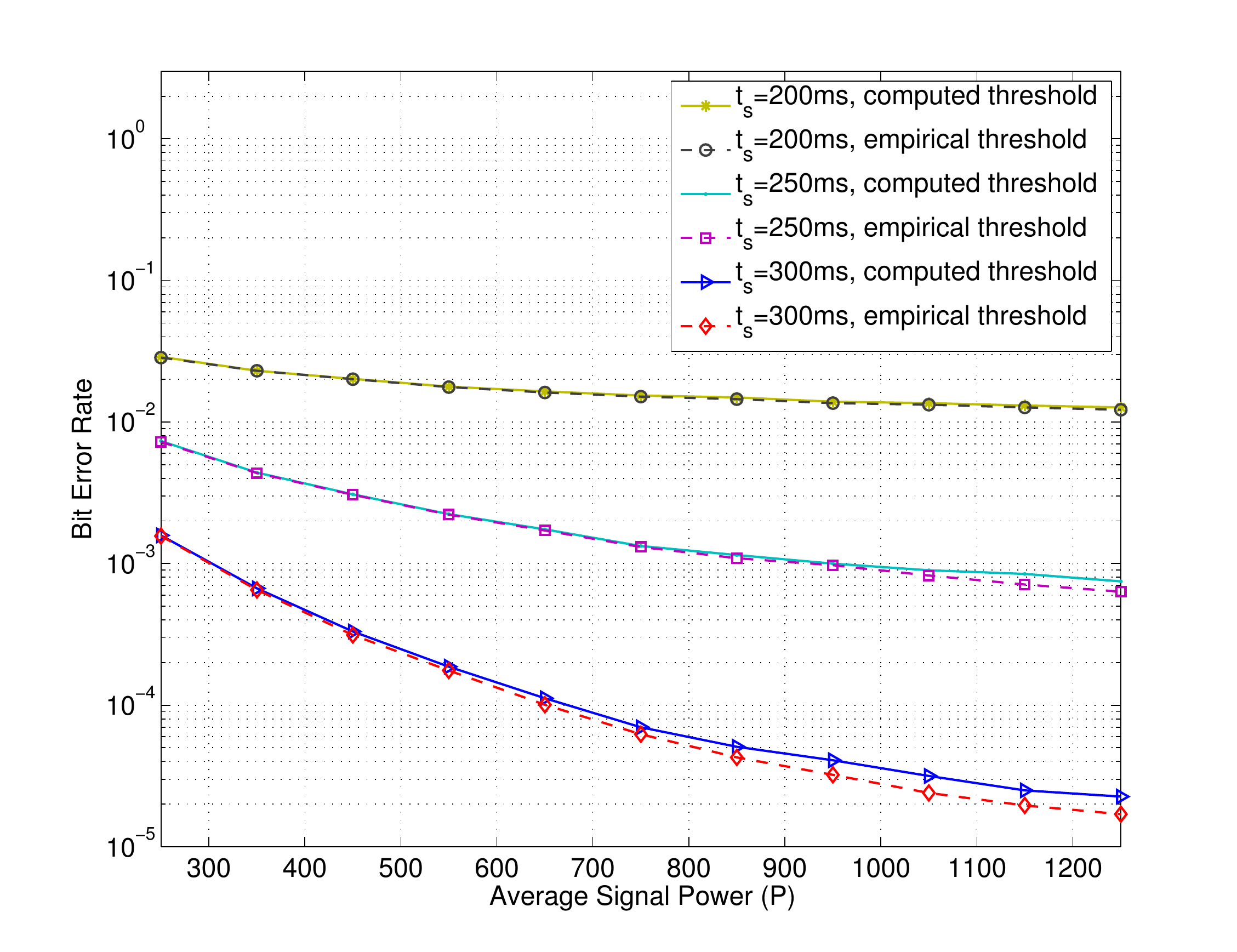}\\
  \caption{Comparision of bit error rates.}
  \label{fig:BER_TH_EMPC}
\end{minipage}%
\begin{minipage}[t]{.5\textwidth}
\centering
 \includegraphics[width=1\columnwidth]{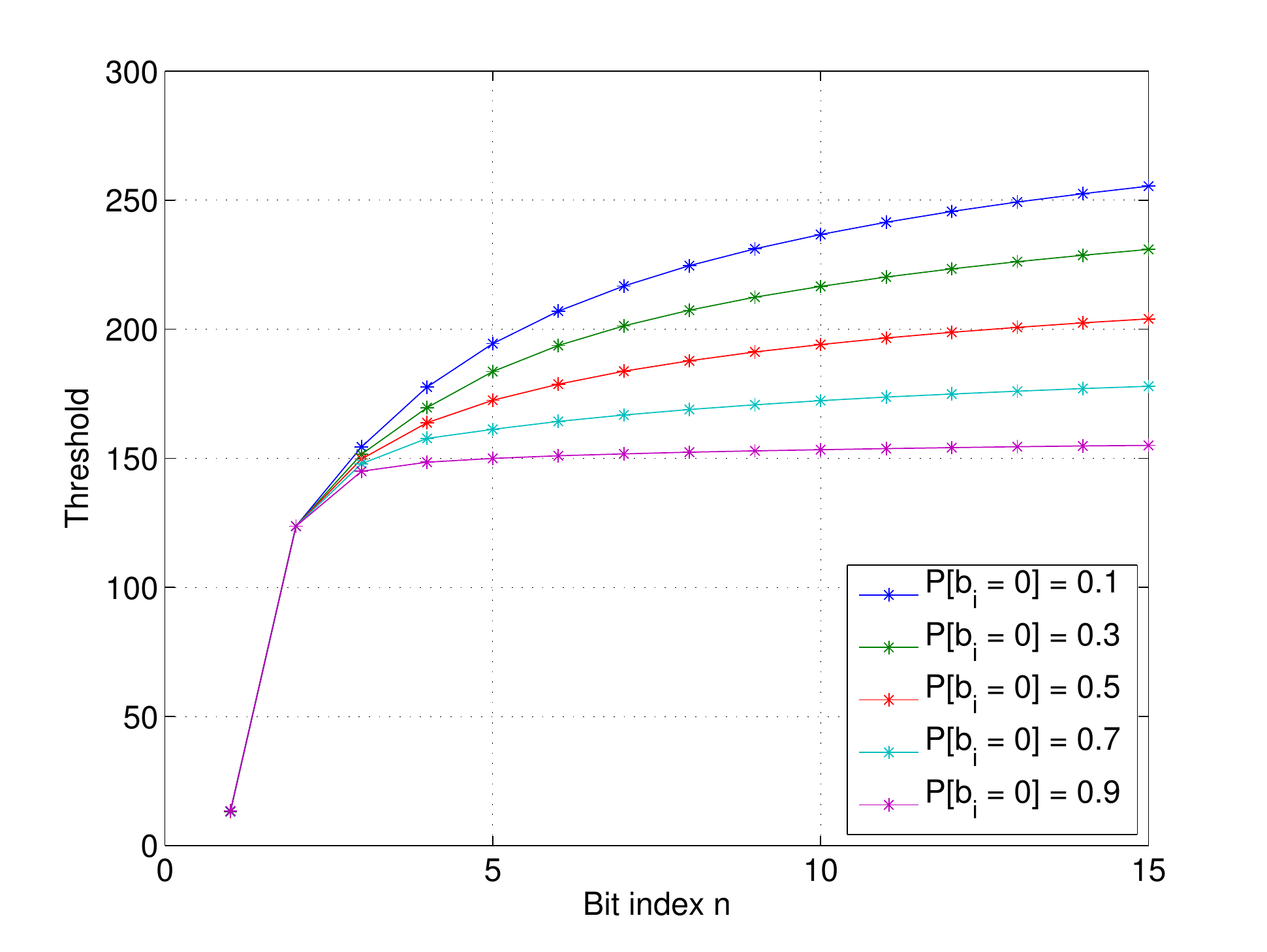}\\
 \caption{Optimal threshold values for different values of $P[b_{i}=0]$, where $t_{s}=200ms$ and $M = 1000$.}
  \label{fig:curves}
\end{minipage}
\end{figure}

Since the distribution parameters depend on the location of each and every bit in the message, each candidate sequence has a unique optimal threshold. In simulations, rather than using pilot symbols, this threshold can be found empirically by trying various number of threshold values and minimizing the Hamming distance $d(\mathbf{\hat{b}}_{1}^{n} ,\mathbf{{b}}_{1}^{n})$ after gathering all the information about molecule counts at the receiver. A performance comparison between the thresholds computed via Algorithm \ref{alg:th} and empirically computed thresholds is given in Figure \ref{fig:BER_TH_EMPC}. Binary sequences of length $10^5$ were used in the simulations, and $\gamma^{*\{10^5\}}$ was applied to every bit without any burn in period.

Empirically found thresholds perform slightly better as expected, since they were found for that particular message, by using the information of the original message $\mathbf{b}_{1}^{n}$. On the other hand, thresholds computed by Algorithm \ref{alg:th} only make use of the the hitting probabilities. Additionally, rate of convergence depends on the value of $P[b_{i}=0]$, since lower $P[b_{i}=0]$ values yield a slower convergence due to the permanently increasing number of accumulating molecules. Curves depending on the numerical values of $P[b_{i}=0]$ are given in Figure \ref{fig:curves}.

\section{Transmitter - based ISI mitigation }

In this section, two transmitter-based ISI mitigation techniques are proposed. The first technique is the molecular transition shift keying (MTSK), which is an energy efficient modulation technique that aims to decrease the detrimental effects of the ISI by utilizing two different molecule types. Second technique employs a power adjustment 
strategy by utilizing the residual molecules from the previous symbols, which is applicable for BCSK, BMoSK, and MTSK.

\subsection{Molecular transition shift keying}

In case of continuous transmission, the first \bt{0} after a large number of consecutive \bt{1}s becomes hard to detect due to the ISI caused by the accumulated molecules in the channel. $\mathbf{b}_{1}^{5} = \{1,1,1,1,0\}$ can be given as an example for such a case. It is hard for the decoder to detect $b_{5}$ due to the ISI caused by $\mathbf{b}_{1}^{4}$. This is the main motivation for MTSK, which aims to distinguish whether the number of received molecules $C_{i}$ in the $i^{th}$ time slot is induced by the ISI due to $\mathbf{b}_{1}^{i-1}$, or a $b_{i}$ being a $1$, by decreasing the amount of ISI due to the residual molecules that belong to previous symbols.

\begin{figure}[!t]
  \centering
  \includegraphics[width=0.7\columnwidth]{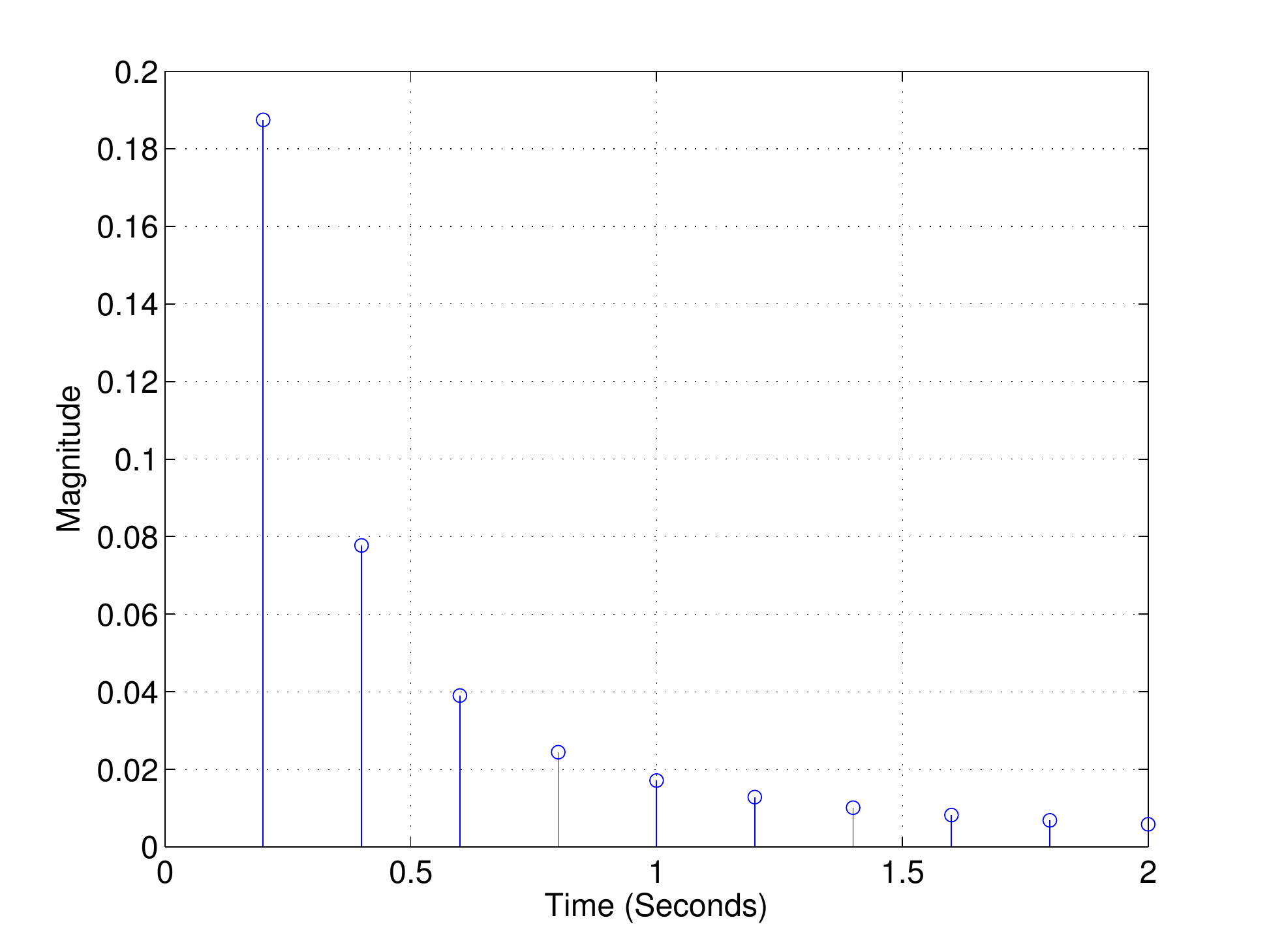}\\
  \caption{First $20$ hitting probabilities, where $t_{s}=200ms$.}
  \label{fig:channelResponse}
\end{figure}

As discussed in Section II, a proper choice of symbol duration allows us to sort the hitting probabilities in a descending order, such as $p_{1}>p_{2}>p_{3}>...$. The amount of decay between consecutive hitting probabilities also decreases decreasingly, which implies that most of the residual molecules that lead to ISI belong to the time slot immediately preceding the current one, and depends on the magnitude of $p_{2}$. Residual molecules from two or more previous time slots which are related to the magnitude of $p_{i}$ for $i \geq 3$ have less significance. To visually illustrate this behavior, the hitting probabilities $p_{k}$ for $k=1,2,...,10$ are given in Figure \ref{fig:channelResponse}.

MTSK can be explained as follows. The \bt{0}s are encoded by the absence of the messenger molecules, and \bt{1}s are encoded by using two different types of molecules, denoted as type-$A$ and type-$B$, of constant number of molecules, $M$, where the choice of the molecule type depends on the value of the following symbol in the message sequence. type-$A$ or type-$B$ molecules are released for the next symbol values of \bt{0} and \bt{1}, respectively. In case of CSK, \bt{1}s are encoded by emitting only type-$A$ (or only type-$B$) molecules, which causes high amounts of ISI observed by \bt{0}s due to the accumulation of the same type of molecules in the channel. On the other hand, in case of MTSK, emitting type-$B$ instead of type-$A$ molecules before each \bt{0} reduces the ISI induced by type-$A$ molecules on each \bt{0}. Similarly, since type-$B$ molecules are only emitted before a \bt{0}, their accumulation in the channel is less than the case where CSK is employed by emitting only type-$B$ molecules. As a result, ISI observed by each \bt{0} is decreased compared to the case where CSK is employed. An example sequence is given in Figure \ref{fig:technique}, where the modulated sequence is delayed for one symbol duration to obtain a causal representation, and $\mathbf{m}_{1}^{n} = \{{m_{1},m_{2},...,m_{n}}\}$ represents the molecule types of the modulated message sequence that will be transmitted through the diffusion channel, where absence of messenger molecules is denoted with an ``$\times$". 

\begin{figure}[h]
\centering
\begin{tabular}{lccccccccccccccccc}
$\mathbf{b}_{1}^{16}$ :  &$0$         & $1$  & $1$  & $1$ & $0$         & $1$  & $0$        & $1$   & $1$ & $0$         & $1$  & $0$          & $0$        & $1$  & $1$ & $0$ \\
$\mathbf{m}_{1}^{16}$ : &$\times$ & $A$ & $A$ & $B$ & $\times$ & $B$ & $\times$ & $A$ & $B$ & $\times$ & $B$ & $\times$ & $\times$ & $A$ & $B$ & $\times$\\
Causal System :&  &$\times$ & $A$ & $A$ & $B$ & $\times$ & $B$ & $\times$ & $A$ & $B$ & $\times$ & $B$ & $\times$ & $\times$ & $A$ & $B$ & $\times$\\ \\
\end{tabular}
  \caption{MTSK modulated binary sequence example.}
  \label{fig:technique}
\end{figure}

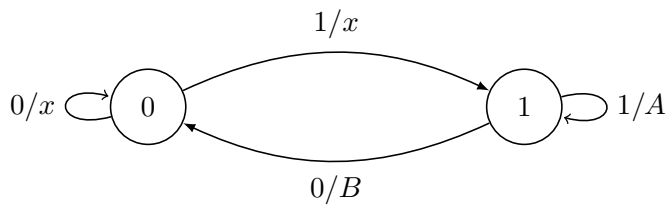
\begin{figure}
\centering
\begin {tikzpicture}[-latex ,auto ,node distance =5 cm and 5cm ,on grid ,
semithick ,
state/.style ={ circle ,top color =white , bottom color = white ,
draw, black , text=black , minimum width =1 cm}]
\node[state] (A) [left] {$0$};
\node[state] (B) [right =of A] {$1$};
\path (A) edge [loop left] node[left] {$0/x$} (A);
\path (B) edge [loop right] node[right] {$1/A$} (B);
\path (A) edge [bend left =25] node[above =0.05 cm] {$1/x$} (B);
\path (B) edge [bend left =25] node[below =0.05 cm] {$0/B$} (A);
\end{tikzpicture}
\caption{State diagram for MTSK encoder.}
\label{fig:markov}
\end{figure}

\noindent This causal system can now be represented by the first-order Markov chain whose state transition diagram is given in Figure \ref{fig:markov}.

Since there are two types of molecules to be sensed in an MTSK modulated signal, a decision should be made based upon the information they jointly possess. Consequently, decoding an MTSK modulated signal requires an optimal choice of decision threshold for both molecule types. Since the receiver is assumed to detect both molecule types independent of each other, $\mathbf{b}_{1}^{n}$ can be treated as two different messages; one modulated by using type-$A$ and the other using type-$B$ molecules, denoted as $\mathbf{b}_{1}^{n}(A)$ and $\mathbf{b}_{1}^{n}(B)$, respectively. $\mathbf{b}_{1}^{n}(A)$ and $\mathbf{b}_{1}^{n}(B)$ are illustrated in Figure \ref{fig:technique2} for the same $\mathbf{b}_{1}^{16}$ in Figure \ref{fig:technique}. Splitting $\mathbf{b}_{1}^{n}$ into $\mathbf{b}_{1}^{n}(A)$ and $\mathbf{b}_{1}^{n}(B)$ allows us to find optimal thresholds for each sequence, which are denoted by $\gamma^{*\{n\}}_{A}$ and $\gamma^{*\{n\}}_{B}$, respectively. 
\begin{figure}[h]
\centering
\begin{tabular}{lcccccccccccccccc}
$\mathbf{b}_{1}^{16}(A)$ : &$0$ & $1$ & $1$ & $0$ & $0$ & $0$ & $0$ & $1$ & $0$ & $0$ & $0$ & $0$ & $0$ & $1$ & $0$ & $0$ \\
$\mathbf{b}_{1}^{16}(B)$ : &$0$ & $0$ & $0$ & $1$ & $0$ & $1$ & $0$ & $0$ & $1$ & $0$ & $1$ & $0$ & $0$ & $0$ & $1$ & $0$ \\
\end{tabular}
  \caption{MTSK encoded binary sequence example.}
  \label{fig:technique2}
\end{figure}

Let $P_{A}[b_{i}=1]$ and $P_{B}[b_{i}=1]$ denote the probability of occurrence for $1$ in $\mathbf{b}_{1}^{n}(A)$ and $\mathbf{b}_{1}^{n}(B)$, respectively. As given in (\ref{eq:bern}), $\mathbf{b}_{1}^{n}(A)$ and $\mathbf{b}_{1}^{n}(B)$ can also be interpreted as $n$ independent Bernoulli trials with probabilities of success $P_{A}[b_{i}=1]$ and $P_{B}[b_{i}=1]$, respectively. Considering that each \bt{1} run of length $r\geq 1$ in $\mathbf{b}_{1}^{n}$ contains exactly one bit encoded by molecule type-$B$, $P_{B}[b_{i}=1]$  and $P_{A}[b_{i}=1]$ can be calculated as

\begin{align}
P_{B}[b_{i}=1] &= P[b_{i}=1](1-P[b_{i}=1]), \label{eq:p1} \\
P_{A}[b_{i}=1] &= P[b_{i}=1] - P[b_{i}=1](1-P[b_{i}=1]) = P[b_{i}=1]^2, 
\label{eq:p2}
\end{align}

\noindent $\gamma^{*\{i\}}_{A}$ and $\gamma^{*\{i\}}_{B}$ can therefore be determined by employing Algorithm \ref{alg:th} using the probabilities in (\ref{eq:p1}) and (\ref{eq:p2}), respectively. 

Let $C_{i}(A)$ and  $C_{i}(B)$ denote the number of type-$A$ and type-$B$ molecules induced at the receiver due to the transmission of $\mathbf{b}_{1}^{i}(A)$ and $\mathbf{b}_{1}^{i}(B)$, respectively. Decision rule for each molecule type can be given~as
\begin{align} 
 C_{i}(A) &\underset{\hat{b}_{i}(A) = 0}{\overset{\hat{b}_{i}(A) = 1}{\gtrless}} \gamma^{*\{i\}}_{A}, \\
 C_{i}(B) &\underset{\hat{b}_{i}(B) = 0}{\overset{\hat{b}_{i}(B) = 1}{\gtrless}} \gamma^{*\{i\}}_{B},
\end{align}
\justifying
\noindent where $\hat{b}_{i}(A)$ and $\hat{b}_{i}(B)$ denote the estimation of ${b}_{i}(A)$ and ${b}_{i}(B)$, respectively. To decide for a \bt{0}, both $\hat{b}_{i}(A) = 0$ and $\hat{b}_{i}(B) = 0$ must be satisfied. On the other hand, if at least one of the number of induced molecules exceeds its corresponding threshold regardless of its molecule type, the decoder decides for a \bt{1}. Consequently, the decision rule for an MTSK encoded binary sequence can be given as
\begin{align}
    \hat{b}_{i} = \begin{cases}
   1 ,& \text{if } C_{i}(A)  > \gamma^{*\{i\}}_{A}  \text{or }  C_{i}(B) > \gamma^{*\{i\}}_{B},  \\
   0 ,& \text{if } C_{i}(A) \leq \gamma^{*\{i\}}_{A} \text{and }  C_{i}(B) \leq \gamma^{*\{i\}}_{B} .
\end{cases}
\end{align}

\begin{example} 
Consider a random binary sequence $\mathbf{b}_{1}^{3} = \{1,1,0\}$, where $P[b_{i}=0]=0.5$, $t_s=200ms$ and $M=100$. To compare the error performance of the BCSK, BMoSK, and MTSK, probability of an erroneous decision for $b_{3}$ will be used, since ISI observed by $b_{3}$ due to previous bits being \bt{1} makes it harder to decode.

Hitting probabilities for these parameters were calculated in Example 1 as $\{p_{1},p_{2}\} = \{0.1875, 0.0777\}$. Using these probabilities, the parameters with their corresponding sequences and threshold values (calculated via Algorithm \ref{alg:th}) are given in Table \ref{table:paramsMod}. Additionally, the decision rule for the BMoSK encoded sequences does not employ a threshold value and it can be expressed as
\begin{align}
 C_{i}(A) &\underset{\hat{b}_{i} = 0}{\overset{\hat{b}_{i} = 1}{\gtrless}} C_{i}(B), 
\end{align}

\noindent since \bt{1} is encoded by molecule type-$A$ and \bt{0} is encoded by molecule type-$B$.

\begin{table}[!t] \centering
\caption{Parameters for BCSK, BMoSK, and MTSK modulation techniques.}
\begin{tabular}{l | l | l | l | l}
Modulation Type      			 & Modulated sequences                                 & Mean (molecules)                         &Variance (molecules$^2$) & Threshold (molecules) \\ \hline 
BCSK                 			 	 & $\mathbf{b}_{1}^{3} = \{1,1,0\}$            &$\mu_{0}^{\{3,3\}} = 11.68$       &$\sigma_{0}^{2\{3,3\}} = 11.92$          		   & $\gamma^{*\{3\}} = 14.64$          \\ \hline
\multirow{2}{*}{\hfill BMoSK} 	 & $\mathbf{b}_{1}^{3}(A) = \{1,1,0\}$              &$\mu_{0}^{\{3,3\}}(A) = 11.68$  &$\sigma_{0}^{2\{3,3\}}(A) = 11.92$                          &$-$           \\
                      				 & $\mathbf{b}_{1}^{3}(B) = \{0,0,1\}$              &$\mu_{1}^{\{3,0\}}(B) = 18.75$  &$\sigma_{1}^{2\{3,0\}}(B) = 16.23$                         &$-$           \\ \hline
\multirow{2}{*}{\hfill MTSK}        & $\mathbf{b}_{1}^{3}(A) = \{1,0,0\}$              &$\mu_{0}^{\{3,1\}}(A) = 3.90$     &$\sigma_{0}^{2\{3,1\}}(A) = 4.75$                         &$\gamma^{*\{3\}}_{A} = 13.76$           \\
                     				 & $\mathbf{b}_{1}^{3}(B) = \{0,1,0\}$              &$\mu_{0}^{\{3,2\}}(B) = 7.77$     &$\sigma_{0}^{2\{3,2\}}(B) = 8.17$                       &$\gamma^{*\{3\}}_{B} = 13.76$          
\end{tabular}
\label{table:paramsMod}
\end{table}

Let $P_{e}^{\text{BCSK}}(b_{i})$, $P_{e}^{\text{BMoSK}}(b_{i})$, and $P_{e}^{\text{MTSK}}(b_{i})$ denote the probabilities of error in the detection of $b_{i}$ for BCSK, BMoSK, and MTSK modulated messages, respectively. These probabilities can be calculated as
\begin{align}
P_{e}^{\text{BCSK}}(b_{3})   &= P[C_{3}>\gamma^{*\{3\}}] = {Q}\left({\frac{\gamma^{*\{3\}} - \mu_{0}^{\{3,3\}}}{\sigma_{0}^{\{3,3\}} }}\right) = 0.1950, \nonumber \\
P_{e}^{\text{BMoSK}}(b_{3}) &= P[C_{3}(B)>C_{3}(A)] = P[C_{3}(B)-C_{3}(A)>0] \nonumber \\
&= {Q}\left({\frac{\mu_{1}^{\{3,0\}}(B)-\mu_{0}^{\{3,3\}}(A)}{\sqrt{(\sigma_{0}^{\{3,3\}}(A))^2+(\sigma_{1}^{\{3,0\}}(B))^2}}}\right) = 0.0913, \nonumber \\
P_{e}^{\text{MTSK}}(b_{3}) &= 1-P[C_{3}(A)<\gamma^{*\{3\}}_{A}]P[C_{3}(B)<\gamma^{*\{3\}}_{B}] \nonumber \\ 
&=1- {Q}\left({\frac{\gamma^{*\{3\}}_{A} - \mu_{0}^{\{3,1\}}(A)}{\sigma_{0}^{\{3,1\}}(A)}}\right){Q}\left({\frac{\gamma^{*\{3\}}_{B}-\mu_{0}^{\{3,2\}}(B)}{\sigma_{0}^{\{3,2\}}(B)}}\right)  = 0.0181. \nonumber
\end{align}
\noindent As a result, $P_{e}^{\text{BCSK}}(b_{3}) > P_{e}^{\text{BMoSK}}(b_{3}) > P_{e}^{\text{MTSK}}(b_{3})$. Note that even though there are only two \bt{1}s before $b_{3}$, differences between error probabilities are noteworthy.
\end{example}

\begin{figure}[!t]
  \centering
  \includegraphics[width=0.7\columnwidth]{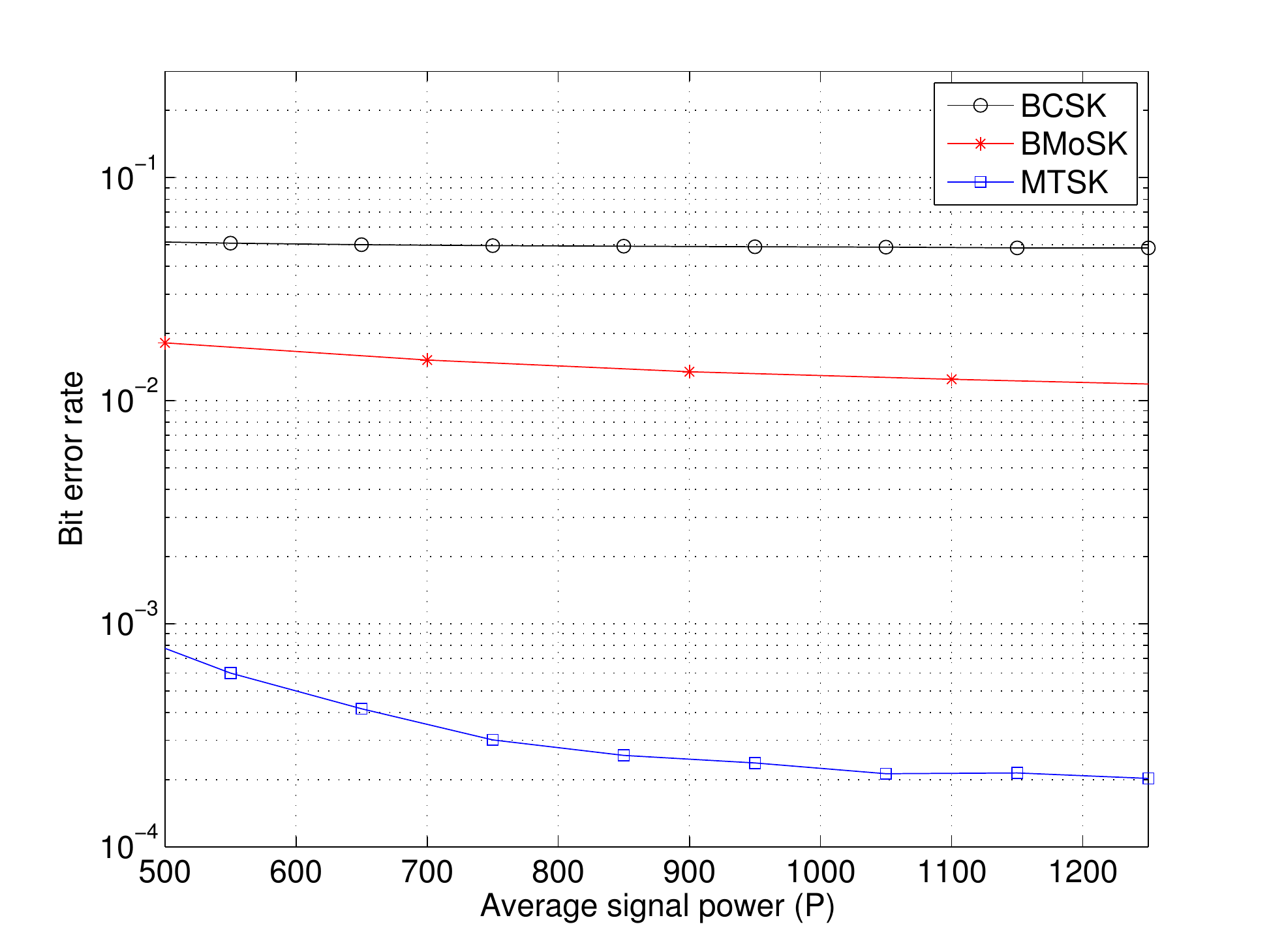}\\
  \caption{BER curves for different modulation techniques, where $t_s=200ms$.}
  \label{fig:BER01}
\end{figure}

In order to compare the error performance of BCSK, BMoSK, and MTSK, average signal power per symbol for these techniques must be defined. For BCSK, where $M_{i}=0$ for $b_{i}=0$, and $M_{i}=M$ for $b_{i}=1$, average power per symbol can be defined as $\bar{P} = M P[b_{i}=1]$. This is also valid for MTSK, since \bt{1}s are encoded with a constant number of $M$ molecules, and \bt{0}s are encoded by the absence of molecules, independent of the molecule type. On the other hand, since BMoSK utilizes $M$ number of molecules for both \bt{0} and \bt{1}, average signal power of BMoSK will be equal to $M$. Error performance of the BCSK, BMoSK, and MTSK modulation techniques are compared via Monte Carlo simulations, and the resulting BER curves are given in Figure \ref{fig:BER01}. $10^4$ realizations were performed, and threshold values computed via Algorithm \ref{alg:th} are used in the simulations. As seen in Figure \ref{fig:BER01}, error rates are significantly decreased when MTSK is employed. By comparison with CSK, employing MTSK increases the system complexity, since it utilizes two different types of molecules instead of one. On the other hand, if utilization of two different molecule types is allowed, one can easily prefer MTSK over BMoSK, since the improvement in the communication quality is very significant.

\subsection{Power Adjustment}

Effects of ISI can both be beneficial (constructive interference) and harmful (destructive interference) to the symbol in question. In BCSK, BMoSK, and MTSK, residual molecules become a source of destructive interference when the intended symbol is a \bt{0}. On the other hand, they may actually be beneficial for consecutive \bt{1}s in a sequence and may be used to support the messenger molecules that will be emitted in the next time slots. With this motivation, BCSK, BMoSK, and MTSK were modified to utilize the residual molecules from the previous symbols, and modified versions are denoted by BCSK-PA, BMoSK-PA, and MTSK-PA, where PA stands for power adjustment. A similar approach on utilizing residual molecules in terms of symbol duration adjustments can be found in~\cite{fekri}. Briefly, in~\cite{fekri}, authors propose a dynamic structure at the receiver side, where they dynamically lengthen the symbol duration as the number of accumulated molecules increase in the channel. This helps to prevent ISI. 

Recall that, in case of a BCSK modulated signal, $M_{i} = 0$ for $b_{i} = 0$ and $M_{i}=M$ for $b_{i}=1$. Since BCSK-PA intends to adjust the signal power considering the effects of constructive ISI,  in case of a BCSK-PA modulated signal, value of $M$ will be adjusted depending on the number of residual molecules in the channel.

Let $E[M_{I}]$ denote the expected value of the number of molecules at the receiver induced by sending $M$ number of molecules for the first \bt{1} in $\mathbf{b}_{1}^{i}$. Since $M_{i}=0$ for $b_{i}=0$, \bt{0}s before the first \bt{1} in $\mathbf{b}_{1}^{i}$ will not effect the number of molecules accumulated in the channel. Relationship between $E[M_{I}]$ and ${M}$ is given as
\begin{align}
E[M_{I}]=p_1 M.
\end{align}

For correct transmission, the threshold at the receiver side should be chosen so that $E[M_{I}]$ number of molecules leads to a symbol decision of \bt{1}. For a sequence containing consecutive \bt{1}s, sending the same amount of molecules for each symbol increases the cumulative number of molecules induced at the receiver side, exceeding $E[M_{I}]$. This will also cause more molecules to accumulate in the channel and become a source of ISI for the following \bt{0}s. On the other hand, for the second and latter symbols, by sending a smaller number of molecules and making use of the residual ones from the previous time slots, $E[M_{I}]$ can still be induced at the receiver and the intended symbol can be decoded correctly. This guarantees the accumulation of fewer molecules in the channel, which, in turn, reduces the amount of ISI for the following symbols. Required number of molecules to maintain $E[M_{I}]$ number of molecules at the receiver after the transmission of the first \bt{1} can be calculated as
\begin{align}
M_{i}=E[M_{I}] - E[M^{R}_{i}],
\label{eq:memo}
\end{align}

\noindent where $E[M^{R}_{i}]$ denotes the expected value of the residual molecules accumulated in the channel due to the transmission of $\mathbf{b}_{1}^{i}$. Since the channel is assumed to be free of messenger molecules before the transmission begins, $E[M^{R}_{i}]$ can be calculated as
\begin{align}
E[M^{R}_{i}] =  \sum\limits_{j=2}^i p_j{M_{i-j+1}}.
\label{eq:res}
\end{align}

Continuously calculating the effects of a large number of symbols from previous time slots is impractical. It is also possible to adjust $M_{i}$ by using a finite memory of length $K$, and rewrite (\ref{eq:res}) as
\begin{align}
E[M^{R}_{i}] =  \sum\limits_{j=2}^K p_j{M_{i-j+1}}.
\label{eq:res2}
\end{align}

In order to apply power adjustment to BMoSK and MTSK modulated signals, expected value of the number of residual molecules must be calculated for both types of molecules, which are denoted by $E[M^{R}_{i}(A)]$ and $E[M^{R}_{i}(B)]$ for type-$A$ and type-$B$ molecules, respectively. Splitting $\mathbf{b}_{1}^{i}$ into $\mathbf{b}_{1}^{i}(A)$ and $\mathbf{b}_{1}^{i}(B)$ as was done in Figure \ref{fig:technique2} allows us to calculate $E[M^{R}_{i}(A)]$ and $E[M^{R}_{i}(B)]$ separately.

The power adjustment technique aims to maintain a constant number of received molecules for a \bt{1}, but by doing so, number of received molecules for a \bt{0} fluctuates depending on $K$, and distorts the monotonically increasing behavior of optimal threshold values. Consequently, $\gamma^{*(i)}$ cannot be calculated for large $i$, and empirically found threshold values are used in the simulations.

\begin{figure}
\centering
\begin{minipage}[t]{.5\textwidth}
  \centering
  \includegraphics[width=1\columnwidth]{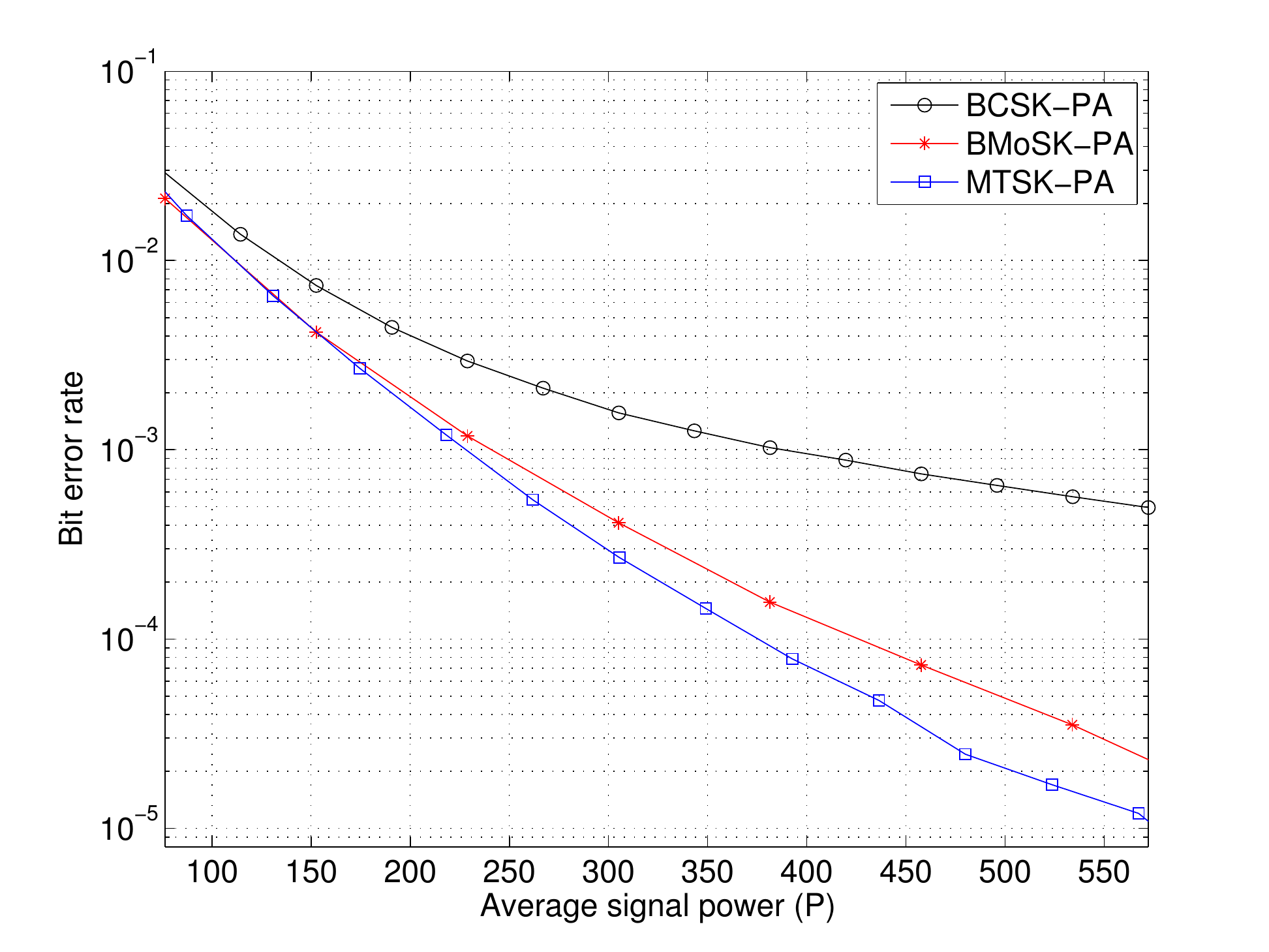}\\
  \caption{BER Curves for different modulation techniques with power adjustment for $K=2$, where $t_s=200ms$ and $P[b_{i}=0]=0.5$.}
  \label{fig:BER02}
\end{minipage}%
\begin{minipage}[t]{.5\textwidth}
  \centering
  \includegraphics[width=1\columnwidth]{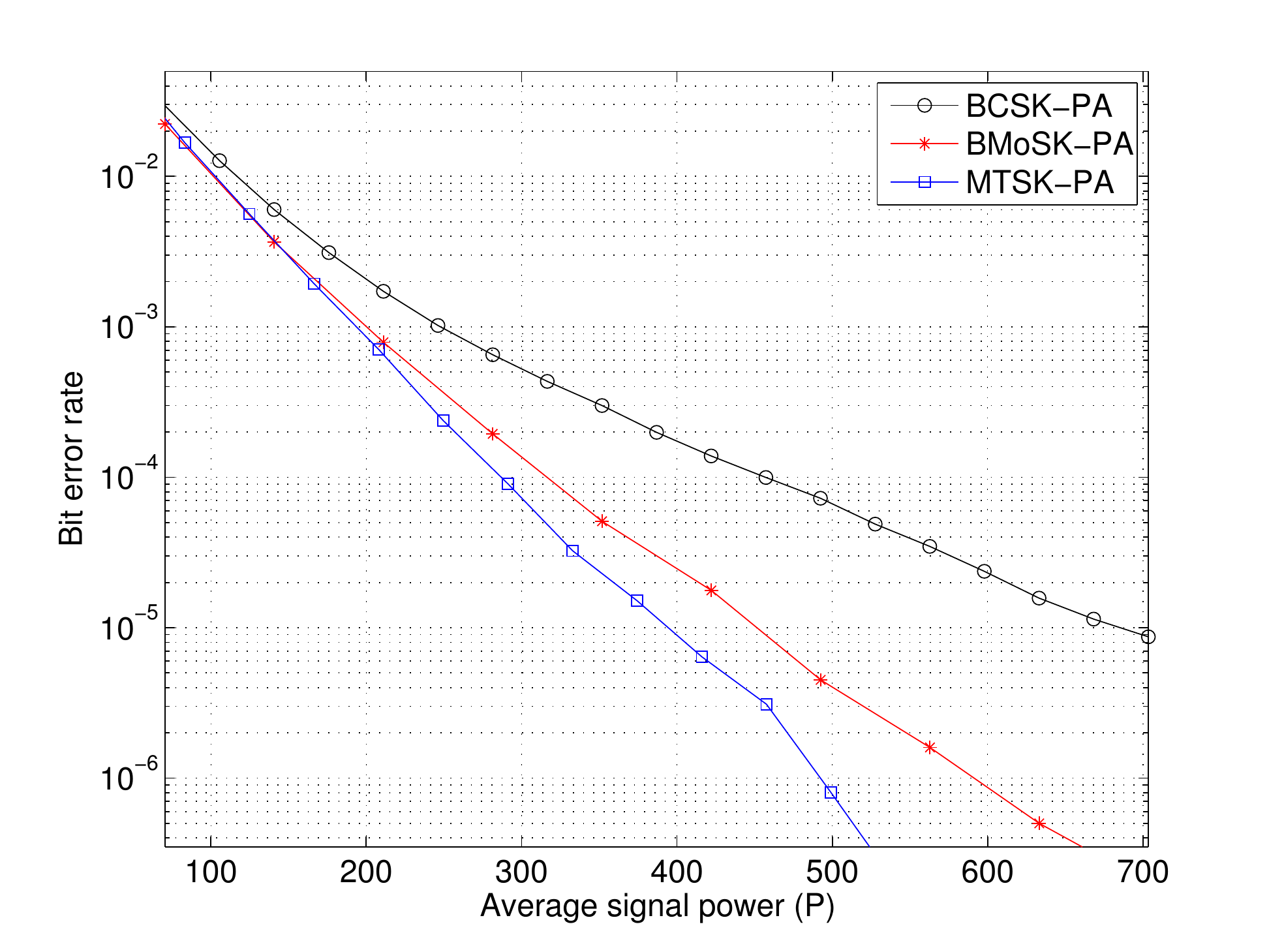}\\
  \caption{BER Curves for different modulation techniques with power adjustment for $K=4$, where $t_s=200ms$ and $P[b_{i}=0]=0.5$.}
  \label{fig:BER03}
\end{minipage}
\end{figure}

Error performance of the BCSK-PA, BMoSK-PA, and MTSK-PA were compared via Monte Carlo simulations and BER curves for $K=2$ and $K=4$ are given in Figure \ref{fig:BER02} and \ref{fig:BER03}, respectively. $15000$ realizations were performed in order to obtain an average. Note that employing power adjustment decreases the error rates for all modulation techniques significantly. Since the effect of $K$ previous bits are considered, as $K$ increases, improvement in the communication quality also increases. On the other hand, increasing $K$ introduces more memory to the system, which results in a trade off between memory length and communication quality.

\section{Receiver – based ISI mitigation}

In this section, we consider a new type of decision-feedback filter for molecular communication with a lower computational complexity than the MMSE equalizer proposed in~\cite{6476072}. Unlike the additive Gaussian noise, the variance of $\C$ is signal dependent, and equalizer tap coefficients must be updated for each sample based on previously detected bits with a feedback mechanism~\cite{6476072}. Each update requires to solve the set of equations given in (30) in~\cite{6476072}, which implies that the computational complexity of the equalizer increases as the number of equalizer taps increases. On the other hand, DFF introduced in this section has a computational complexity independent of the number of filter taps, but requires a larger number of memory elements in order to achieve the same BER with the MMSE filter. 

As mentioned in Section III, the number of candidate sequences $\mathbf{d}_{1}^{\{i,j\}}$ and $\mathbf{d}_{0}^{\{i,j\}}$ increases with powers of $2$ as $i$ increases. On the other hand, when $\hat{\mathbf{b}}_{1}^{i-1}$ is available to receiver, there are only two candidate sequences conditioned on \bt{0} and \bt{1}, which are $\mathbf{d}_{0}^{\{i,j\}} = \{\hat{\mathbf{b}}_{1}^{i-1},0\}$ and $\mathbf{d}_{1}^{\{i,j\}}=\{\hat{\mathbf{b}}_{1}^{i-1},1\}$, respectively. $\mu_{1}^{\{i,j\}}$, $\mu_{0}^{\{i,j\}}$, $\sigma_{0}^{\{i,j\}}$, and $\sigma_{1}^{\{i,j\}}$ can be calculated using (\ref{eq:mean01}) and (\ref{eq:std02}), and these parameters can be used to calculate $\gamma^{*\{i,j\}}$ via solving the quadratic equation given in (\ref{eq:quad}). Consequently, by storing previously estimated bits at the receiver, $\gamma^{*\{i,j\}}$ for each $i$ can be calculated in a signal dependent manner, assuming that the decisions are correct. 

\begin{figure}
\centering
\begin{minipage}[t]{.5\textwidth}
\centering
  \includegraphics[width=1\columnwidth]{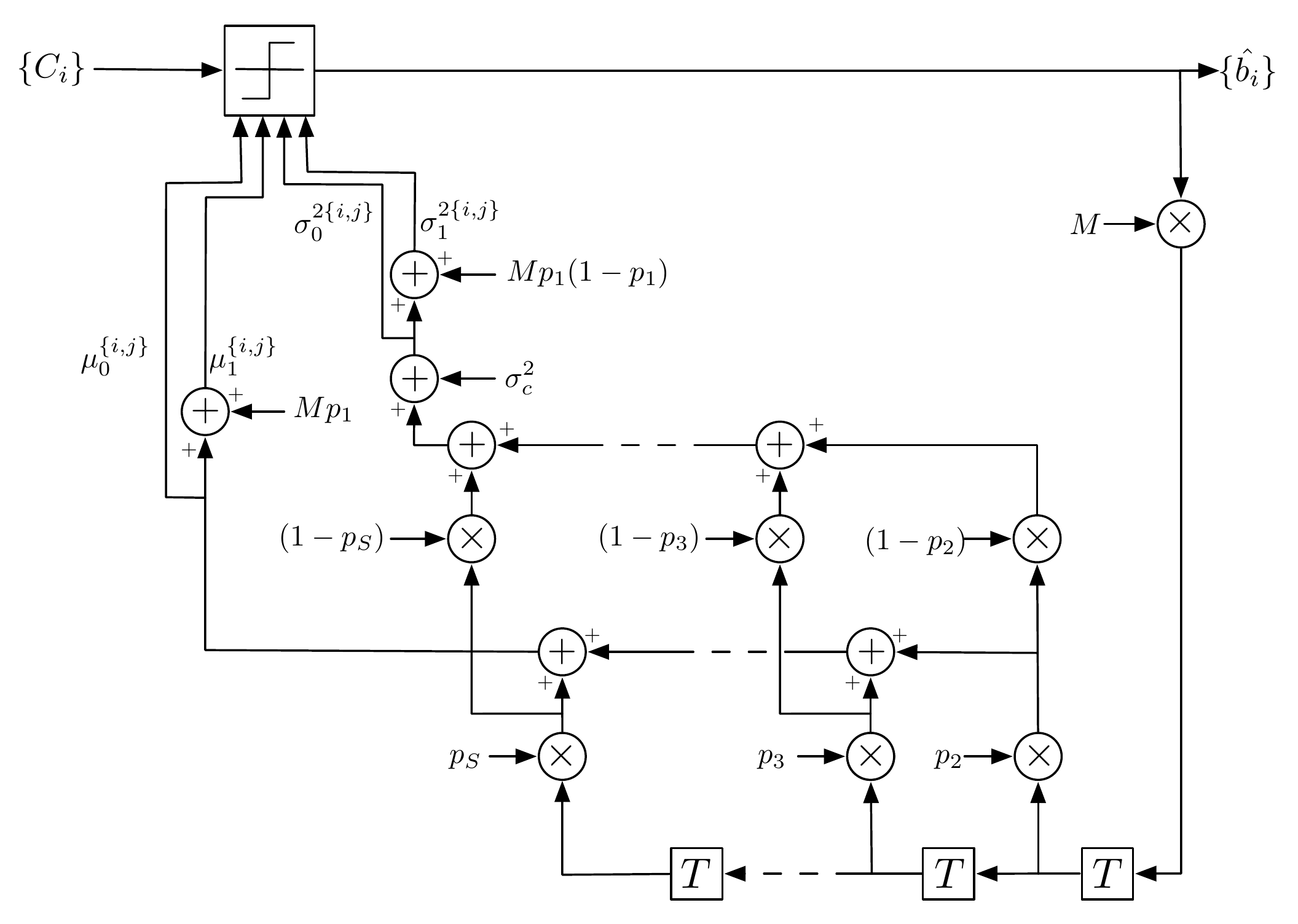}\\
  \caption{Block diagram of decision feedback filter.}
  \label{fig:DFF}
\end{minipage}%
\begin{minipage}[t]{.5\textwidth}
\centering
  \includegraphics[width=1\columnwidth]{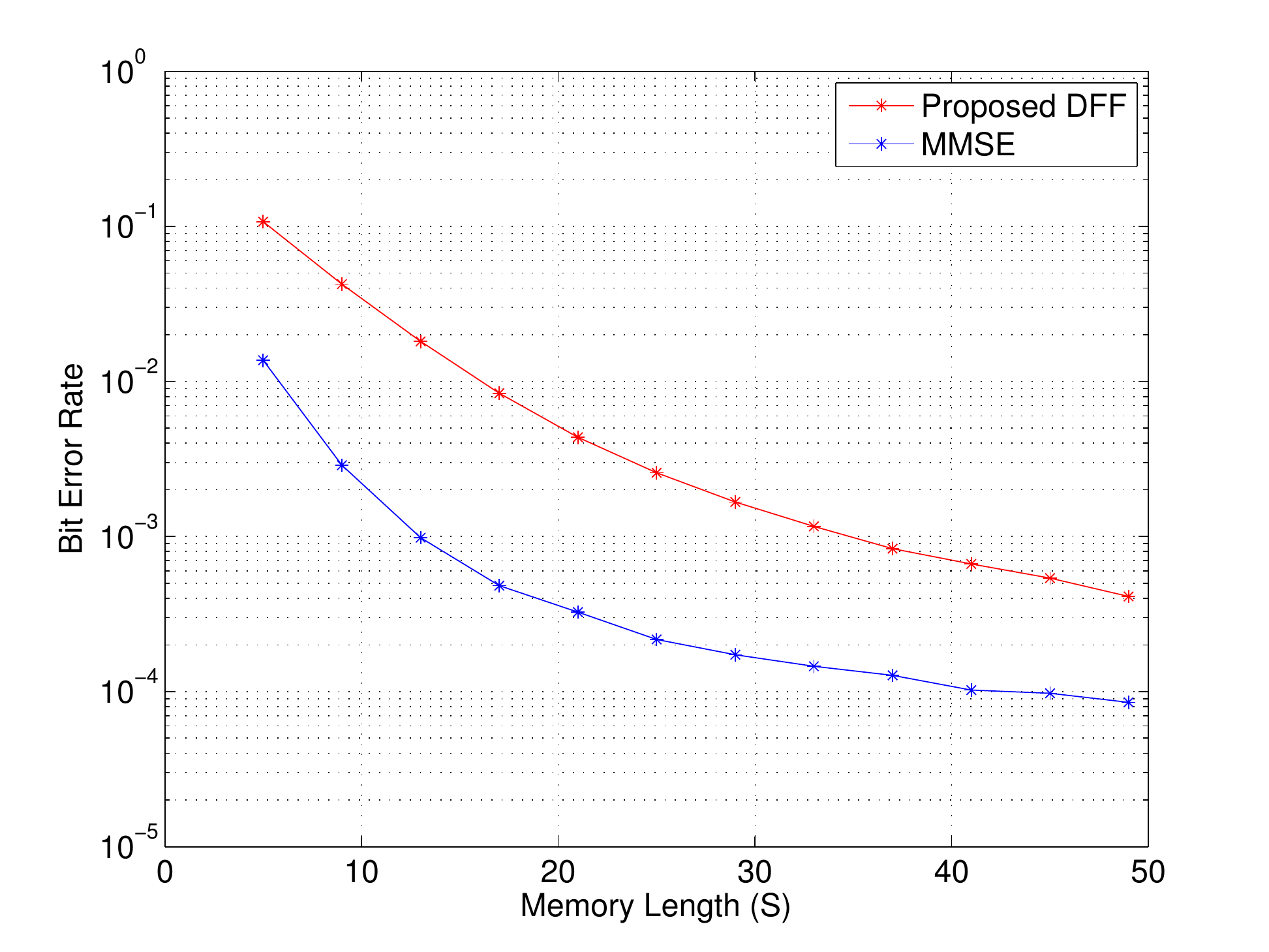}\\
  \caption[]{BER Curves for DFF and MMSE equalizer\footnotemark, where $M=500$ and $t_s=200ms$.}
  \label{fig:DFF_ber}
\end{minipage}
\end{figure}

In case of continuous transmission, storing $\hat{\mathbf{b}}_{1}^{i-1}$ for each $i$ requires infinite memory, and is hence impractical. Rewriting  distribution parameters for a finite memory receiver yields to
%
\begin{align}
\label{eq:paramsDFF}
\mu_{0}^{\{i,j\}} &= M\sum_{k=2}^{S}p_{k}\,\hat{b}_{i-k+1}, \\
\mu_{1}^{\{i,j\}} &= M\sum_{k=2}^{S}p_{k}\,\hat{b}_{i-k+1} \,+ Mp_{1}, \\
\sigma_{0}^{2\{i,j\}} &=  \sigma_{c}^2+ M\sum_{k=2}^{S}p_{k}\,(1-p_{k})\,\hat{b}_{i-k+1}, \\
\sigma_{1}^{2\{i,j\}} &=  \sigma_{c}^2+ M\sum_{k=2}^{S}p_{k}\,(1-p_{k})\,\hat{b}_{i-k+1} + p_{1}\,(1-p_{1}),
\end{align}
\noindent where $S$ denotes the length of the receiver memory. Block diagram of this DFF is given in Figure \ref{fig:DFF}. Note that the use of erroneously detected $\mathbf{\hat{b}}_1^{i-1}$ may cause error propagation, which may decrease the performance of the DFF.

Error performance of DFF and the MMSE equalizer proposed in~\cite{6476072} are compared via Monte Carlo simulations and bit error rates with respect to memory lengths  are given in Figure \ref{fig:DFF_ber}. $500$ realizations were performed in order to obtain an average.


\footnotetext{MMSE filter proposed in~\cite{6476072} stores $S=2K-1$ previously detected bits to calculate $K$ equalizer taps.}

As seen in Figure \ref{fig:DFF_ber}, in order for the DFF and the MMSE equalizer to perform at a bit error rate of around $10^{-3}$, DFF must have $S=35$, whereas $S=13$ is sufficient for MMSE equalizer. However, if the computation of the distribution parameters are ignored, computational complexity of MMSE equalizer is at the order of $\mathcal{O}(S^3)$, whereas computational complexity of DFF is equal to $\mathcal{O}(1)$, since it only requires to solve the quadratic equation given in (\ref{eq:quad}). 
\section{Results and discussion}

In this paper, transmitter and receiver-side energy efficient ISI mitigation techniques were proposed for MCvD in terms of modulation, filtering, and signal power adjustment. To work on modulation techniques for a real time communication scenario, decision threshold for detection had to be determined prior to the information transmission, which was a problem never addressed in the literature before. An analytical method that minimizes the overall error rate when the system parameters are known was proposed to determine the optimal decision threshold for each symbol. Since the number of operations performed for calculating the optimal threshold increases for the latter symbols in the sequence, LMS regression is applied to the first $20$ threshold values, using a function that resembles the dynamics of the diffusion channel. By doing so, optimal thresholds can be calculated for each symbol regardless of the sequence length. These threshold values are compared with the empirically found threshold values via Monte Carlo simulations, and are verified to be optimal in the sense of minimizing the overall bit error rate. It is concluded that, as long as $\alpha$, $\beta$, and $\kappa$ in (\ref{eq:lms}) are known, different thresholding strategies can be applied depending on the system constraints. Resolving the thresholding problem is the pre-requisite step which allows us to propose new ISI mitigation techniques.

The first transmitter-based solution proposed for ISI mitigation is a novel modulation technique, titled MTSK, which utilizes the use of multiple molecule types in order to increase the data rate via suppressing the negative impact of the ISI on communication quality. It was shown via Monte Carlo simulations that MTSK decreases the bit error rates significantly, and outperforms the two most common modulation techniques in the literature, which are BCSK and BMoSK. Furthermore, as the second transmitter-based solution, a power adjustment technique, which utilizes the residual molecules in the channel, is proposed in order to enhance the energy efficiency. Error performance of CSK-PA, MoSK-PA, and MTSK-PA were compared via Monte Carlo simulations, and it was shown that the power adjustment technique decreases the ISI, hence the bit error rate for a fixed signal power for all modulation techniques, significantly. Furthermore, a trade off is observed between memory length ($K$) employed in PA and communication quality.

The receiver-based solution proposed for the energy efficiency problem was to employ a simpler decoder in terms of computational complexity, titled decision feedback filter. DFF calculates the optimum threshold value for the symbol in question and updates the decision threshold for each sample by using the previously estimated bits. DFF was compared with the MMSE equalizer proposed in~\cite{6476072} in terms of bit error rate, memory length, and computational complexity via Monte Carlo simulations. It was concluded that DFF requires more memory to reach the same error rate as that of the MMSE equalizer, but since calculating the optimal threshold value has computational complexity at the order of $\mathcal{O}(1)$, DFF becomes more advantageous when energy efficiency is a priority. 

\section*{Acknowledgment}
The work of B. Tepekule, A. E. Pusane, and T. Tugcu was in part supported by the Scientific and Technical Research Council of Turkey (TUBITAK) under grant number 112E011, Bogazici University Research Fund (BAP7436), and the State Planning Organization (DPT) of the Republic of Turkey under the project TAM (2007K120610). The work of H. B. Yilmaz and C.-B. Chae was in part supported by the MSIP (Ministry of Science, ICT \& Future Planning), Korea, under the ``IT Consilience Creative Program" (NIPA-2014-H0201-14-1002) supervised by the NIPA (National IT Industry Promotion Agency) and by the Basic Science Research Program (2014R1A1A1002186) funded by the Ministry of Science, ICT and Future Planning (MSIP), Korea, through the National Research Foundation of Korea.


\bibliographystyle{IEEEtran}
\bibliography{IEEEabrv,nanoHuge}

\end{document}